\documentclass[letterpaper,groupedaddress,showpacs,nofootinbib,showkeys,preprint,12pt]{revtex4-1}
\usepackage{graphicx,amscd,amsmath,amssymb}

\usepackage{graphicx}
\usepackage{multirow}
\usepackage{xcolor}
\bibstyle{apsrev}

\begin{document}

\title{Tunable Contact Resistance in Transition Metal Dichalcogenide Lateral Heterojunctions}

\author{Adam Pfeifle and Marcelo A. Kuroda}
\email{mkuroda@auburn.edu}
\affiliation{Department of Physics, 
Auburn University, Auburn, AL 36849, USA}

\begin{abstract}

Contact resistance of semiconducting transition metal dichalcogenides has been shown to decrease in lateral heterojunctions formed with their metallic phases but its origins remain elusive. Here we combine first principles and quantum transport calculations to rationalize the contact resistance of these structures in terms of phase, composition (WTe$_2$, MoTe$_2$, WSe$_2$, and MoSe$_2$), and length of the channel. We find that charge injection in metallic 1T'-WTe$_2$/1T'-MoTe$_2$ junctions is nearly ideal as electrode Bloch states remain delocalized through the channel. Mixtures of 1T' selenides and tellurides depart from this scenario due to the momentum mismatch between states in the lead and channel. In semiconducting channels, the large Schottky barriers degrade the electrical contacts.  Around band edges, contact resistance values are about an order of magnitude lower than those obtained experimentally suggesting that doping and phase-engineering could be employed to overcome this issue. We predict that transport regime in these junctions shifts from thermionic emission to tunneling for channels shorter than 3~nm~at room temperature. We also discuss the presence of states at the metal/semiconductor interfaces. By underpinning mechanisms to control the contact resistance in heterogeneous two-dimensional materials, this work proves valuable towards the development of devices suitable for optoelectronics and phase-change materials applications.
\end{abstract}
\date{\today}


\maketitle

Atomically thin crystals \cite{novoselov2005a} have become unique platforms to novel physical phenomena unseen in the bulk \cite{gupta2015}.  These two-dimensional (2D) materials -- such as graphene, hexagonal boron nitride, transition metal chalcogenides, and phosphorene among others --  have attracted great attention due to their diverse electrical, mechanical, thermal, and optical properties \cite{Novoselov2004, Novoselov2005, Radisavljevic2011,Chen2018,Han2008,Hu2012}.  Within this growing class of materials, much interest in the pursuit of nanoscale device applications has been given to the group VI transition metal dichalcogenides (TMDs) with the stoichiometric formula $M\!X_2$, where $M$ is Mo or W, and $X$ is a member of the chalcogen family (S, Se or Te) \cite{chhowalla2013}. Group VI TMDs are mostly semiconducting with band gaps of up to 1.9~eV \cite{wang2012} making them potential candidates for ultrathin, flexible low-power electronics and valleytronics \cite{cao2012,mak2012}. In addition, a special feature of this group of 2D crystals is the small energy difference between the semiconducting (2H) phase and the metallic distorted octahedral structure (1T') \cite{duerloo2013, ChemReview2015}. This polymorphism sets them apart from other 2D materials as exceptional systems with reduced dimensionality exhibiting metal-insulator transitions under ambient conditions \cite{duerloo2016} and, thereby, promising systems for phase-change memory devices \cite{sebastian2018,raoux2014}.

Contact resistance in TMDs stands as a key issue for the realization of optoelectronic device applications based on these materials \cite{yoon2011,Allain2014}. Extensive efforts have focused on the optimization and control of the Schottky barriers between semiconducting channel material and metal electrodes \cite{leonard2011, das2013a, chuang2014, kumar2015, Kim2017}. The contact resistance of lateral heterojunctions \cite{mahjouri2015, li2015a, Cho2015} has significantly diminished with respect to the values obtained through direct metal deposition onto 2H TMD nanosheets \cite{Chen2013, Popov2012}.  These promising  approaches rely on inducing TMDs into their metallic phase \cite{kappera2014,Retamal2018} to produce stable heterophase junctions via chemical vapor deposition \cite{Sung2017, Yang2017, Leong2018} or laser-induced phase patterning \cite{Cho2015}. Despite efforts to date, the fundamental mechanisms governing contact resistance in these heterojunctions, its fundamental limits, and means to control carrier injection via phase-engineering are still missing.

In this work, we characterize from the atomistic viewpoint the in-plane contact resistance of lateral heterojunctions based on group VI TMDs using first principles calculations within the density functional theory (DFT). Specifically, we combine electronic structure and quantum transport calculations to identify the underlying physical principles limiting charge carrier injection into metallic and semiconducting TMD channels by accounting for effects stemming from the composition, phase, and length.  In metal-metal junctions, transmission is nearly ideal for telluride-telluride systems, where the delocalized Bloch states near the Fermi level overlap across a larger range of reciprocal space than in telluride-selenide systems.

In metal-semiconductor junctions, Schottky barriers emerge for both electrons and holes that significantly diminish transmission. Our estimates of the contact resistance are comparable to those obtained experimentally, but yet an order of magnitude higher than the lowest attainable values (5~$\Omega \cdot \mu$m) due to a reduced overlap between states near the Fermi level.  These results suggest that doping could surmount this issue by both lowering the barriers and enhancing coupling between states in the electrode and channel.  For short channel lengths, carrier injection is dominated by tunneling or coupling through metal induced gap states (MIGS) formed at the interfaces. We discuss the dispersion of these states based on the geometry of the interface and their implications to transport. Our quantitative analysis predicts the length scale below which tunneling dominates electrical transport over thermionic emission based on operating conditions. By establishing the mechanisms that determine the contact resistance in TMD junctions, this work may prove valuable to the advancement of low-power electronic devices based on 2D materials.

\section{Methodology}

We describe the contact resistance in lateral heterostructures formed by TMDs with different compositions and phases using first principles calculations within DFT.  Under ambient conditions, monolayers of all group VI TMDs (except WTe$_2$) exist in the 2H structure \cite{Rao2017}; WTe$_2$ is found to be stable in the asymmetric 1T' structure \cite{Lee2015}.  Therefore, we mainly focus on heterogeneous systems where the electrodes are formed by four orthorhombic unit cells (uc)  of 1T'-WTe$_2$ on each side of the channel.  Channels with various compositions $M\!X_2$, where $M=$ W or Mo, and $X=$ Se or Te, are considered in both their 1T' and 2H phases (see Figure~\ref{fig:system}a and Figure~\ref{fig:system}b).  The channel length of heterojunctions in this work range from 2 to 6 uc ($\sim$12-36~\AA). These systems resemble the sharp interfaces grown experimentally where either the metal or the chalcogen elements are substituted \cite{mahjouri2015,li2015a}.  Our TMD channels are selected based on the small formation energy difference between these two polymorphs \cite{duerloo2016}. 

The geometries of the systems are obtained via structural relaxation of supercells where the lattice constant of the system $a_{0}$ is set to that of the 1T'-WTe$_2$ electrode ($a_0=3.45$~\AA) while the channel length is allowed to vary.  The atoms located away from the interface are kept fixed preserving the shape of the bulk lead; the remaining atomic positions are fully relaxed until Hellmann-Feynman forces on each atom were lower than 0.01~eV/\AA. The supercell assumes periodic boundary conditions with at least 12 \AA~vacuum space between planes.  We employ the Perdew-Burke-Ernzerhof (PBE) \cite{Perdew1997} parameterization of the exchange-correlation functional while accounting for van der Waals (vdW) dispersive forces. \cite{Thonhauser2015,Thonhauser2007,Berland2015,Langreth2009}.  Calculations presented here do not include spin-orbit coupling.

 Description of core electrons is performed with ultrasoft pseudopotentials \cite{Dion2004,Roman2009} using energy cutoffs of 40 Ry and 400 Ry for the wave function and density, respectively.  The self-consistent density is determined using  $12\!\times\!1$~$k$-point grid. We compute the ballistic conductance using the PWCOND implementation of the Quantum Espresso software package \cite{Choi1999,Giannozzi2009,Smogunov2004}. A 96~$k_x$-point grid is employed in the transport calculations to sample the 1D-BZ and compute the $k_{x}$-resolved transmission $T(k_{x},E)$.

\section{Results}

\begin{figure}[htpb]
\centering{
\includegraphics[width= 3.3in]{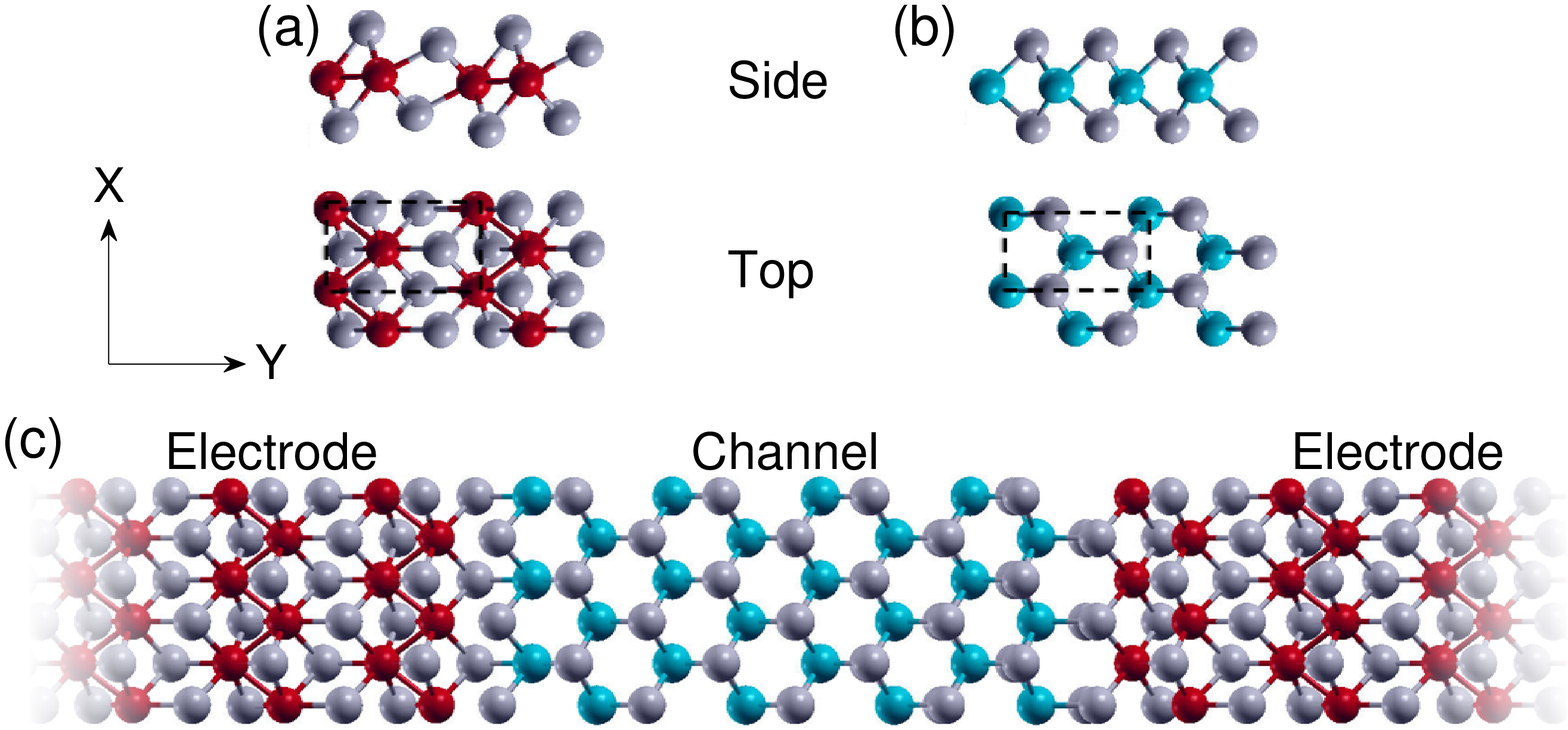}
\includegraphics[width= 3.2in]{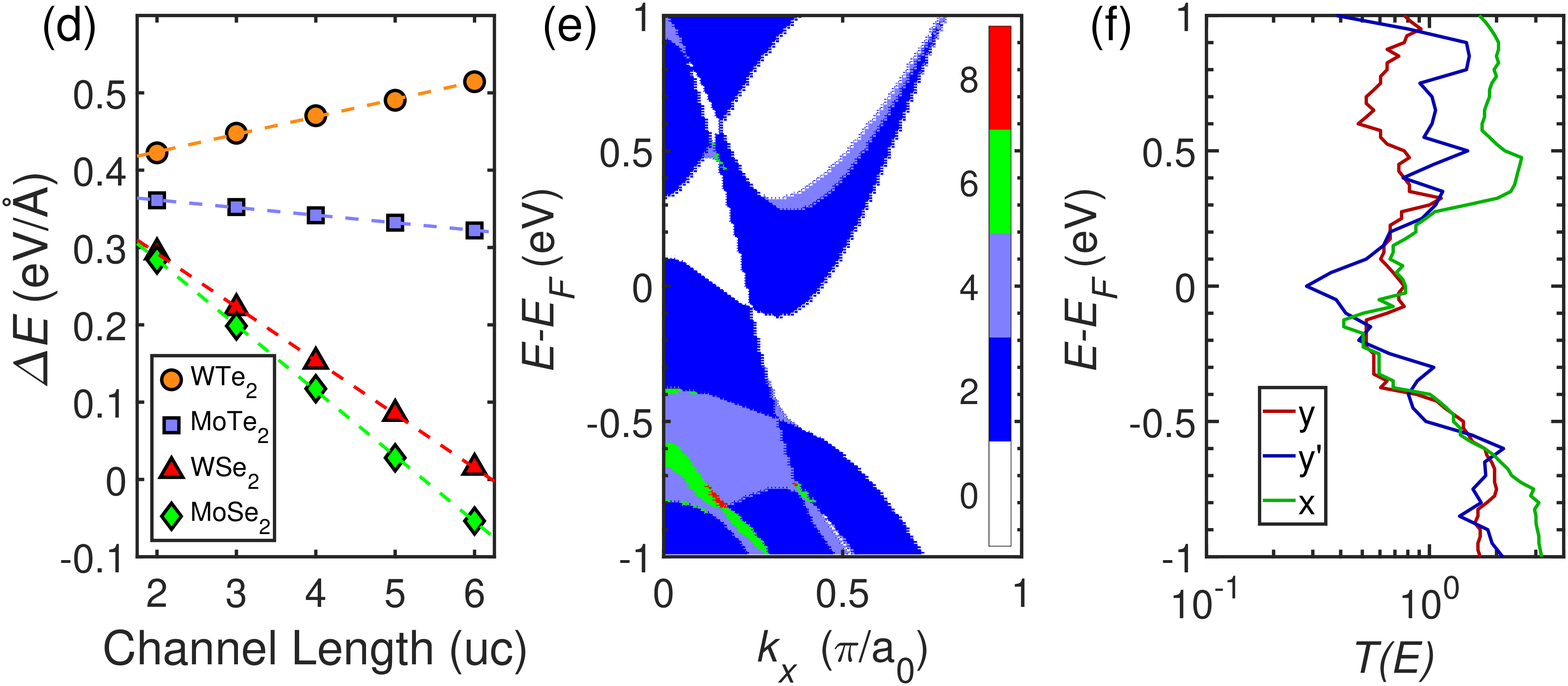}
\caption{Top and side view of (a) 1T' and (b) 2H $M\!X_2$ bulk crystals.  The dashed lines denote the orthorhombic unit cell. (c) Example heterojunction formed by 1T'-WTe$_2$ electrodes and 4~uc-long channel formed with 2H-MoTe$_2$. (d) Energy difference between systems with 2H and 1T' channels ($\Delta\!E = E_{2H} - E_{1T'}$) normalized by the supercell width and various compositions: WTe$_2$, MoTe$_2$, WSe$_2$, and MoSe$_2$, as obtained from DFT calculations. Dashed lines correspond to $MX_2$ formation energy. (e) $k_x$-resolved transmission probability $t_{e\!f\!f}(k_x,E)$ (Eq.~\ref{eq:teff}) for bulk 1T'-WTe$_2$ along the $\hat{y}$ direction. (f) Comparison of transmission $T(E)$ of bulk 1T'-WTe$_2$ for transport along directions at 0$^\circ$, 31.2$^\circ$, and 90$^\circ$ angles with respect to the $\hat{y}$-axis.
}\label{fig:system}}
\end{figure}

We first determine the ground state energy of heterojunctions with 1T'-WTe$_2$ leads  where the channel composition is maintained, but its phase and length are varied. The obtained energy difference between these two phases normalized by the supercell width, namely $\Delta E \equiv E_{2H}-E_{1T'}$, for WTe$_2$,  MoTe$_2$,  WSe$_2$, and  MoSe$_2$ are presented in Figure~\ref{fig:system}d. For short channel lengths, the 1T' configuration appears to be more stable than the 2H (i.e.~$\Delta E  > 0$) inheriting the ground-state crystal structure of the WTe$_2$ leads.

As channel length increases, the energy difference $\Delta E$ decreases linearly, due to the fact that the 2H structure is the most stable among the different phases of group VI TMDs, save WTe$_2$ which shows an increase \cite{duerloo2013}. Indeed, the slopes are consistent with the bulk formation energy difference per $M\!X_2$ formula for these phases (S1). For the selenides, the 2H channel becomes more stable than their 1T' phase counterparts after approximately 6 uc. Extrapolation of the data presented in Figure~\ref{fig:system}d serves to estimate the grain boundary energy between the 1T' contact and the 2H channels, which yields similar results ($\sim$0.4 eV/\AA) irrespectively of the TMD channel composition.  Further tunability of this 2H-1T' transition \cite{Zhang2016}, relevant to applications in phase change materials \cite{sebastian2018},  may be attained by alloying the TMD channels \cite{Park2015}. 

In these crystalline junctions, carrier injection finds its origins in different mechanisms such as direct tunneling, thermionic emission of either electrons or holes \cite{Sze1981,Padovani1966}, or coupling via MIGS \cite{Picozzi2000}. We primarily focus on analyzing the contact resistance based on the atomistic details of these heteroepitaxial structures. Other mechanisms limiting transport, such as the electron-phonon coupling, scattering with impurities/defects and interactions with substrates, should also be included when describing larger systems. 

As junctions considered in this work have channel lengths of up to 6 uc~($\sim$36~\AA), we assume that transport is ballistic and characterize the dependence of contact resistance on composition, phase, and length of the channel. In the ballistic regime within the Landauer formalism \cite{buttiker1982,datta1995}, the junction resistivity $r_c$ can be determined  using \cite{Carlo1994}:

\begin{equation}
r_c = \frac{a_{0}}{G_0\, T(E_F)} \label{eq:resist},
\end{equation}
where $G_0=e^2/h \approx 7.75\times10^{-5}$~S is the conductance quantum, $a_{0}$ is the supercell's width, and $T(E_F)$ is the average transmission probability at the Fermi level ($E_F$).  
In Eq.~\ref{eq:resist}, the transmission probability $T(E_F)$ reads:

\begin{equation}
T(E_F) =\! -\!  \int \!dE \left[\frac{a_{0}}{2\pi} \int_{\tiny\mbox{1D-BZ}} \,dk_{\perp}\, t_{e\!f\!f}(k_\perp,E)\right]\frac{df}{dE} \label{eq:barT},
\end{equation}
where $f(E)$ and $t_{e\!f\!f}(k_\perp,E)$ are the Fermi-Dirac distribution and the $k_\perp$-resolved (transverse to the transport direction) transmission probability averaged over the one-dimensional Brillouin zone (1D-BZ). The probability $ t_{e\!f\!f}(k_\perp,E)$ reads: 
\begin{equation} 
t_{e\!f\!f}(k_\perp,E) =  \sum_{i,j} t_{i,j}(k_{\perp},E) \label{eq:teff}
\end{equation}
and accounts for contributions from all modes with crystal momentum $k_\perp$ originating in the source electrode (labeled by $i$) coupling to  modes in the drain (labeled by $j$).

For any material forming the leads, the number of available states propagating along a given direction \cite{kuroda2011} establishes the upper limit in transmission seen in Eq.~\ref{eq:resist}. We quantify the case of 1T'-WTe$_2$, employed as electrodes in our heterojunctions, to set a baseline for the optimal transmission through our systems, where the coupling of all modes in the leads to the channel is ideal. A profile of the bulk momentum-resolved transmission $t_{e\!f\!f}(k_\perp,E)$ through 1T'-WTe$_2$ leads for transport along the $\hat{y}$ direction is provided in Figure~\ref{fig:system}e. Energies with overlapping bands yield higher transmission values, which are multiples of 2 because of spin-degeneracy. To gauge possible dependencies on the transport direction, we compute the transmission of bulk 1T'-WTe$_2$ (Eq.~\ref{eq:barT}) along three different orientations: $\hat{x}$, $\hat{y}$, and $\hat{y}'$, set 31.2$^\circ$ past $\hat{y}$.

We find that the ideal contact resistance $r_c$ for 1T'-WTe$_2$ is approximately 5 to 15 $\Omega\cdot \mu$m based on Eq. ~\ref{eq:resist} using transmission values between 0.3 to 0.85 at the Fermi level as shown in Figure~\ref{fig:system}f. These values are well below currently attained values in TMD junctions which exceed 100~$\Omega\cdot \mu$m \cite{kappera2014,Cho2015,Chuang2016}. Moreover, variations in the maximum achievable transmission (of up to a factor of 2) are small compared to effects of phase and composition analyzed here. Hence, for the remainder of this study, we compute transport along the $\hat{y}$-axis as the lattice mismatch between 1T'-WTe$_2$ (metal contact) and other group VI TMDs (channel) is smallest in the $\hat{x}$ direction. The small changes in band structures of 1T'-TMDs suggest that our results are weakly affected by the inclusion of spin-orbit coupling in metal-metal heterojunctions (S2).

\begin{figure}[htpb]
\centering
\includegraphics[width= 3.25in]{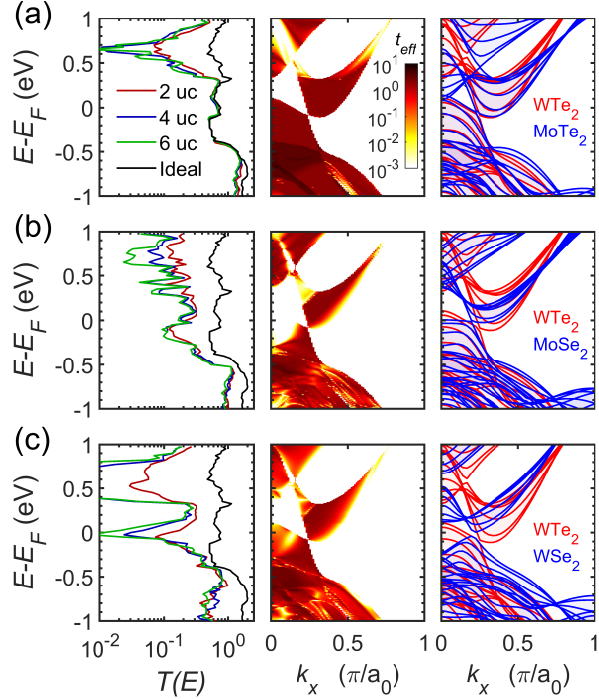}
\caption{(a) MoTe$_2$ channel (b) MoSe$_2$ channel (c) WSe$_2$ channel: (left) Average transmission $T(E)$ for 1T'-1T' junctions formed with WTe$_2$ contacts for each composition with 2 (red), 4 (blue), and 6 (green) uc long channels. As a reference, the ideal transmission attainable with bulk 1T'-WTe$_2$ is also plotted (black).  (middle) $k_x$-resolved transmission probability $T(k_x,E$) of channel materials in the 1T' phase. (right) Electronic bands of bulk channel materials (blue) overlaid bulk 1T'-WTe$_2$ (red) along the 1-D Brillouin zone path. Shaded regions indicate areas with commensurate band structure that correlate with regions of increased transmission. 
}\label{fig:1t_trans}
\end{figure}	

We next analyze the transmission probability in metal-metal junctions, whose channels are formed by MoTe$_2$, MoSe$_2$, and WSe$_2$ in their 1T' phase.  Their average transmissions $T(E)$  (Eq.~\ref{eq:barT}) with channel lengths ranging from 2 to 6 uc are plotted in Figure~\ref{fig:1t_trans}.  Outcomes of these junctions show varying behaviors with respect to the ideal case. Remarkably, MoTe$_2$ channels exhibit almost optimal conductance near the Fermi level although it decreases a few orders of magnitude at energies 0.5 eV above the Fermi energy. In contrast, heterojunctions containing MoSe$_2$ or WSe$_2$ channels show reductions in transmission of an order of magnitude or larger than the ideal case in the vicinity of the Fermi level. We find that transmission between 1T'-WTe$_2$ and 1T'-Mo$X_2$ ($X=$ Te or Se) is weakly dependent on channel length near the Fermi level. In spite of their metallic nature, channels of 1T'-WSe$_2$ show vanishing transmissions with increasing channel length near the Fermi level, as well as at energies above 0.5~eV. 

Insights on the origins of transmission reductions in metal-metal heterojunctions are obtained from the momentum-resolved transmission profiles [$t_{e\!f\!f}(k_x,E)$] for each composition and the corresponding band structures of the bulk crystals. The profiles for 4 uc-long channels reveal that only MoTe$_2$ preserves high transmission values for states with energies below 0.25~eV. In the other cases, coupling between modes is drastically attenuated leading to the overall reduction in transmission observed in the first column of Figure~\ref{fig:1t_trans}.   We also note that, given both the channel and the lead, an improved predictor of the upper limit in the transmission may be obtained from the comparison of the availability of modes in the bulk of both components (S3).

These attenuations in transmission that increase the junction contact resistance emerge due to the momentum mismatch between modes in the electrode and channel \cite{kuroda2011}. To exemplify this, in Figure~\ref{fig:1t_trans} we overlay the electronic band structures of the bulk for channel materials along $k_x$-paths parallel to the $\Gamma$-X onto that of the electrode (1T'-WTe$_2$). The case of MoTe$_2$ evinces little reduction in $T(E)$ in $k_x$-$E$ regions where the electrode and channel bands overlap, allowing the nearly ideal transmission of electrons. In contrast, energy ranges where band structures exhibit with small or no coincidence  (e.g.~the case of WTe$_2$/WSe$_2$ near the Fermi level) produce the large reduction in transmission, as transport becomes governed by tunneling or a very small region in momentum space.

\begin{figure}[htpb]
\centering{
\includegraphics[width= 3.25in]{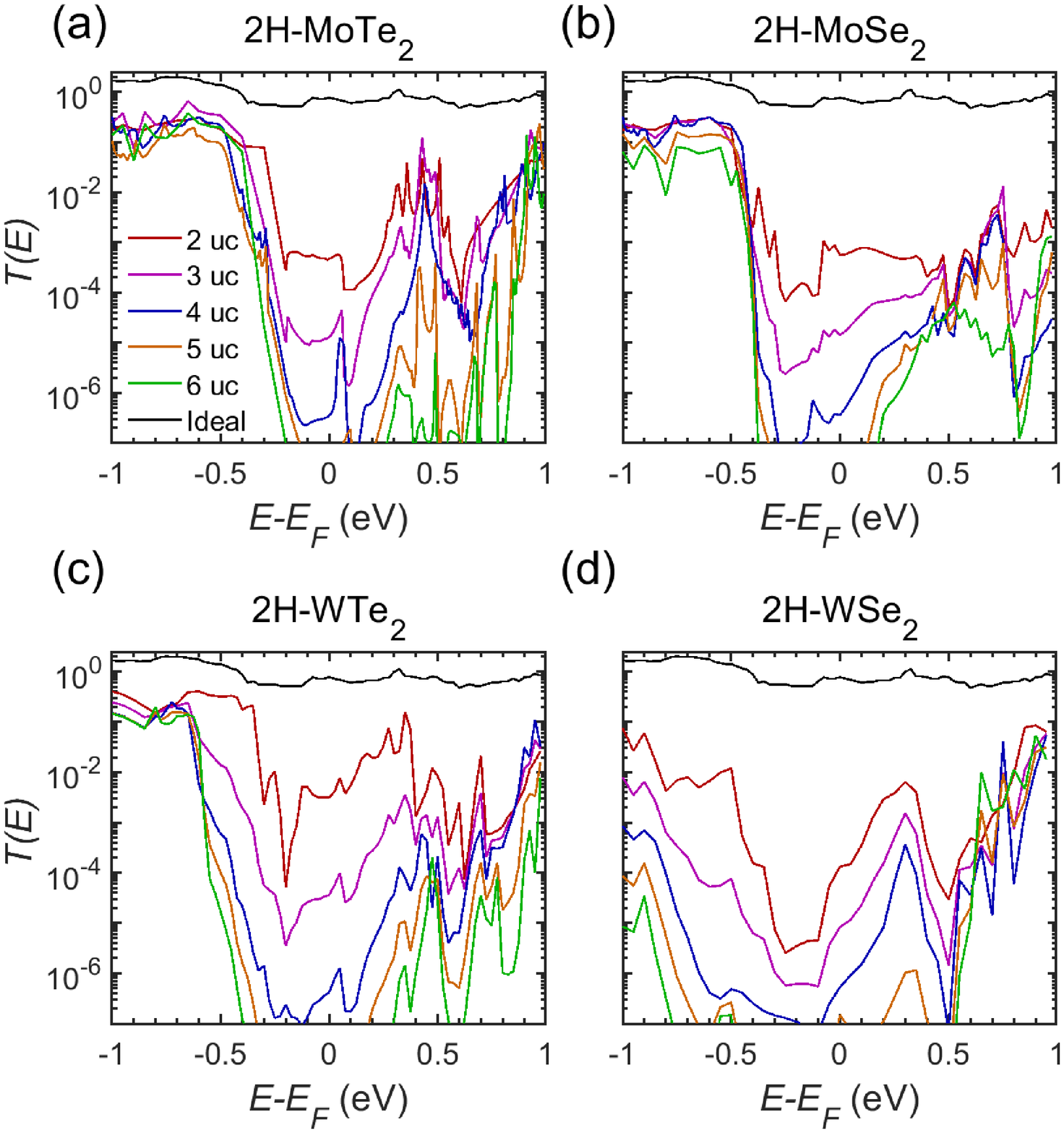}\\
\includegraphics[width= 3.15in]{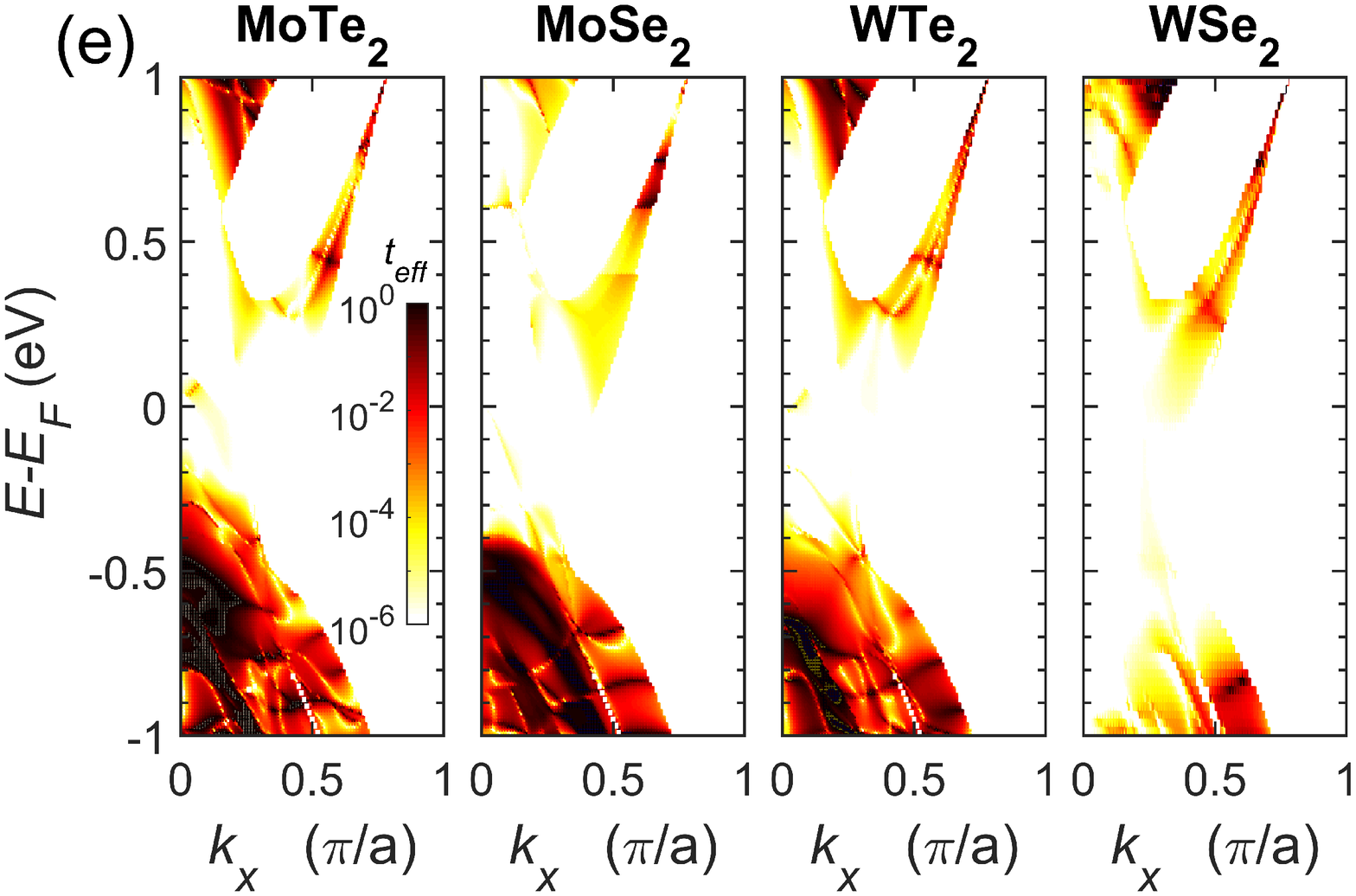}
\caption{Average transmission $T(E)$ for junctions formed  with 1T'-WTe$_2$ leads and different semiconducting 2H channels: (a) MoTe$_2$, (b) MoSe$_2$, (c) WTe$_2$, and (d) WSe$_2$, respectively.  The channel lengths plotted here correspond to 2 (red), 3 (purple), 4 (blue), 5 (orange) and 6 (green) uc.  As a reference, the maximum attainable transmission for the WTe$_2$ lead is also plotted (black). (e) Momentum-resolved transmission $t_{e\!f\!f}(k_x,E)$ profiles for 4 unit cell long 2H channels  with 1T'-WTe$_2$ leads.  From left to right: MoTe$_2$, MoSe$_2$, WTe$_2$ and WSe$_2$.
}\label{fig:2h_trans}}
\end{figure}

Using a similar approach to that of metal-metal junctions, we find that systems with 2H channels behave differently due to their semiconducting character. These calculations include the same group VI TMDs as in the 1T' channels with the addition of 2H-WTe$_2$ and channel lengths ranging from 2 to 6~uc. The average transmissions $T(E)$ of these systems, provided in Figure~\ref{fig:2h_trans}, diminishes several orders of magnitude with respect to the maximum attainable close to the Fermi level as band edges reside far from it. According to these DFT results, Schottky barriers of varying heights are formed at these interfaces yielding nonohmic contacts.

Determining the relative band alignment between electrode and semiconducting channels is challenging as channel lengths considered in this work are shorter than band bending characteristic distances in these materials \cite{zhang2014a} and could change in the presence of defects or dopants \cite{Tosun2016,Xu2017}.  Here the position of the band edges is inferred from band structure decomposition of the full system (S4).  In the structures considered here we find that band offsets -- presumably arising from dipoles at the interface -- are different for selenides and tellurides.  These shifts are of +0.1~eV and -0.1~eV for the 2H $M$Te$_2$ and $M$Se$_2$ systems, respectively. Valence band edges reside at least 0.5~eV below the Fermi level; conduction bands, on the other hand, are located about 0.6~eV above or higher.   

Analysis of $k_x$-resolved transmission probabilities $t_{e\!f\!f}(k_x,E$) in Figure~\ref{fig:2h_trans} reveals a drastic attenuation of modes within this energy range which leads to the overall reduction in transmission. Transmissions $T(E)$ become nearly independent of channel length at energies far from the Fermi level (except for WSe$_2$). This behavior, similiar to that found in most 1T'-1T' junctions, arises from the coupling between states in the channel and electrode. Small deviations from this behavior (e.g. -1 eV in Figure~\ref{fig:2h_trans}a) can be attributed to the charge transfer between the channel and electrode, as well as effects from quantum confinement. Electron injection declines significantly at these energy levels due to only a small portion of the $k_x$-space region contributing to transport (see Figure~\ref{fig:system}e).

For carrier energies at the band edges (Figure~\ref{fig:2h_trans}), the highest average transmission values through the 6 uc-long 2H channels are of the order of 10$^{-1}$,  yielding contact resistances $r_c$ of about 50~$\,\Omega\cdot$$\mu$m for holes and slightly higher for electrons. These resistance values are smaller than those obtained experimentally in 1T/2H-phase engineered TMD lateral heterojunctions (100-1000~$\,\Omega\cdot$$\mu$m) \cite{kappera2014,Retamal2018,Leong2018,Cho2015}, indicating that the charge carrier injection may still allow for improvements discussed next. 

For lateral metal-semiconductor TMD junctions, transmission is influenced by the barrier heights and the coupling between electrodes and channel.  Thus, phase-engineering and doping may be employed to reduce the contact resistance by exploiting defects \cite{Tosun2016}, impurities \cite{Xu2017}, or the formation of epitaxial lateral heterojunctions \cite{mahjouri2015,li2015a,Leong2018}. Aside from lowering the barriers for electrons or holes, this mechanism can effectively raise transmission by enlarging the Fermi level density of states coupling to the semiconducting channel, at the expense of reducing transmission at other energy levels. This analysis may be carried out from the overlap of the band structures for the given epitaxy as presented in the supplemental information (S3).  Interestingly, splitting due to spin-orbit coupling does not yield significant differences in the estimates of our systems because of the momentum mismatch \cite{kuroda2011} between modes in regions where the splitting is significant.

For short length channels, the main transport mechanism is direct tunneling and length dependence of transmission rates can be approximated by $T(E) \propto e^{-2{\kappa_{e\!f\!f}}L}$, where $\kappa_{e\!f\!f}$ and  $L$ are the tunneling rate and the channel length, respectively. We estimate $\kappa_{e\!f\!f}$ for each system using the zero-bias transmission probability of each TMD channel as a function of length (S5). We find decay rates ranging between 0.22 and 0.33 \AA$^{-1}$ for 2H TMDs, which are consistent with calculations based on the bulk complex band structures (S6).

As the length of 2H channels increases, carrier injection becomes governed by thermionic emission. For this mechanism, transmission is sensitively dependent on the Boltzmann factor $T(E) \propto e^{-\frac{E_b}{k_BT}}$, where $E_b$ is the height of the Schottky barrier, $k_B$ is the Boltzmann constant, and $T$ is temperature. We define the crossover length $\ell$ between transport regimes as the distance when tunneling exponential decay equals the Boltzmann factor for thermionic emission [$\ell\equiv E_b/(2k_BT\kappa_{e\!f\!f})$]. We find that the crossover lengths $\ell$ are of the order of 3~nm for MoTe$_2$,  MoSe$_2$, and WTe$_2$, and slightly larger (5~nm) for WSe$_2$ due to its wider band gap. We point out that these estimates would likely be reduced upon with the inclusion of spin-orbit provided that those modes indeed couple to the electrodes (S2/S6). Moreover, smaller Schottky barriers from interfacial dipoles and MIGS could enhance conduction at or near the Fermi (see for example Figure~\ref{fig:2h_trans}a).  The energy location and extension of these evanescent states depend on the 2H-1T' interface, as discussed later.

\begin{figure*}[htpb]
\centering{
\includegraphics[width= 5.in]{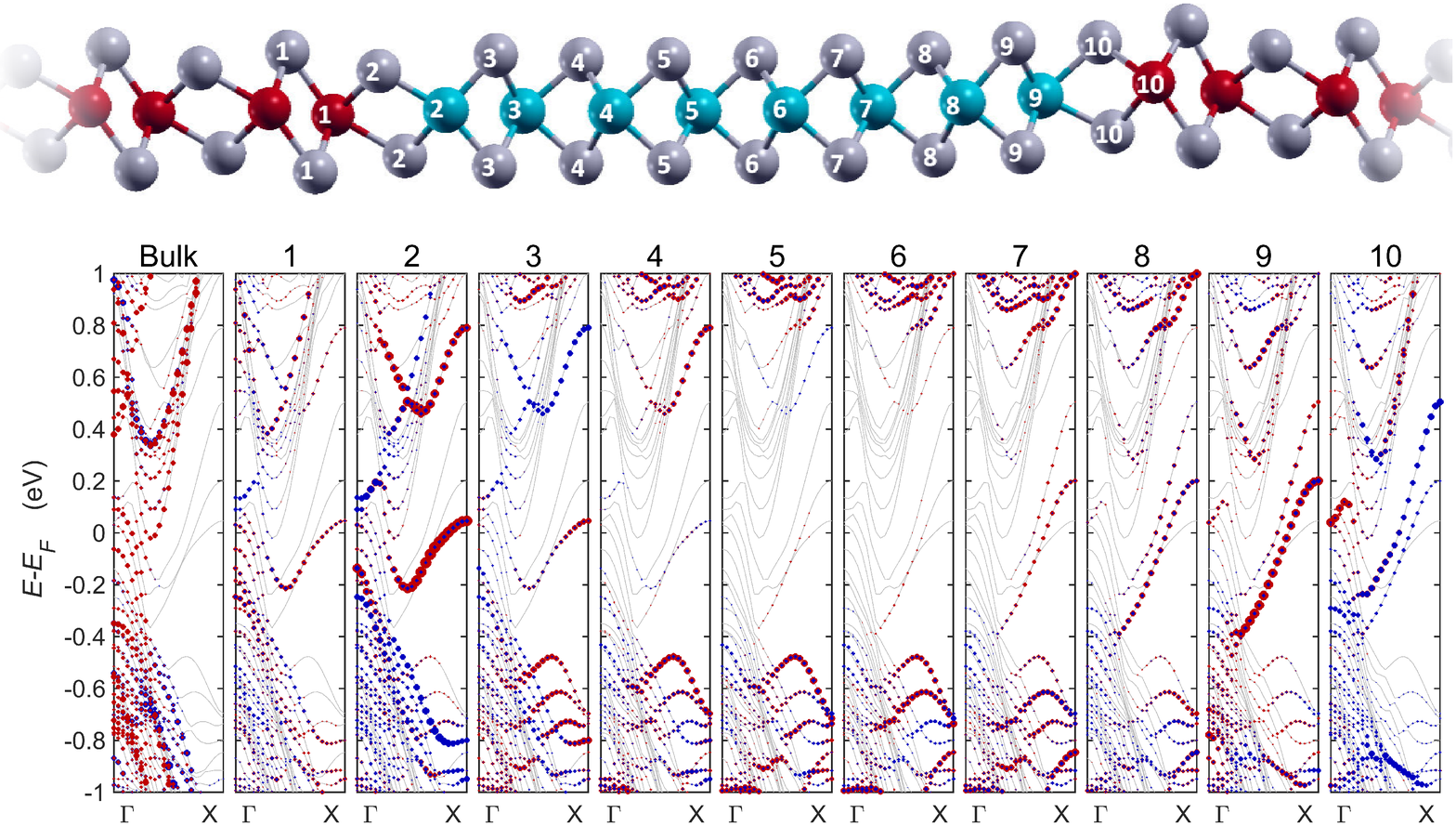}
\caption{Geometric and electronic structure of a heterojunction with 1T'-WTe$_2$ leads and 4 uc-long 2H-MoTe$_2$ channel. (Top) Side view of junction. Atoms in the junction are assigned to different regions in the channel and at the interface with the leads as labeled. (Bottom) Bloch states of the supercell are projected onto localized atomic orbitals of atoms (red - transition metal, blue - chalcogen) located in the different regions along the channel as indicated in above. (left) metal lead; (right) supercell split into 10 contributions per $M\!X_2$ along channel. 
Interface states can be observed in the second and ninth segment, corresponding with the beginning and end of the channel.
}\label{fig:pdos}}
\end{figure*}
	
To further understand transport through the different 2H channels, we parse the semiconducting channel and its interface in different regions parallel to the interface.  In Figure~\ref{fig:pdos} we plot the projections onto localized atomic orbitals (LAOs) of atoms in each region as labeled in the diagram of the system above. This band structure decomposition allows us to locate conduction and valence bands throughout the channel, MIGS, and the momentum match/mismatch between states in the leads  (1T'-WTe$_2$) and those of the $M\!X_2$ for the 2H-MoTe$_2$ channel. For instance, Bloch states reside either 0.5~eV below or 0.8~eV above the Fermi level in the middle of the channel, suggesting the absence of ohmic contacts in these heterojunctions. Similar results have been previously reported for MoS$_2$ \cite{marian2016}.  A comparison between the dispersion of states in the middle of the channel and the bulk electrode (left panel) shows limited crystal momentum overlap ($k_x$), especially for the conduction bands.  As a result, a poor carrier injection of electrons in the MoTe$_2$ channel is obtained with large barrier heights ($E-E_F>0.8$~eV). 

At intermediate energies, we observe the presence of MIGS at the edges of the channel (regions 1-3 and 8-10 in Figure~\ref{fig:pdos}).  These states create paths for carrier injection via tunneling at different energy levels  ($E-E_F\sim -0.1$~eV or 0.5~eV) that vanish as the length of the channel increases (Figure~\ref{fig:2h_trans}).  On both sides, one of the interface states \cite{Ugeda2018,Wang2019} crosses the Fermi level while the other remains above it.  In addition, the asymmetric termination of the junction yields different $k_x$ dispersion for these interface states \cite{Chen2017}. As these states may be relevant to the nonlocal transport through interface states \cite{roth2009,Ovando2019}, it is important to analyze their features.

\begin{figure}[htpb]
\centering{
\includegraphics[width= 3.5in]{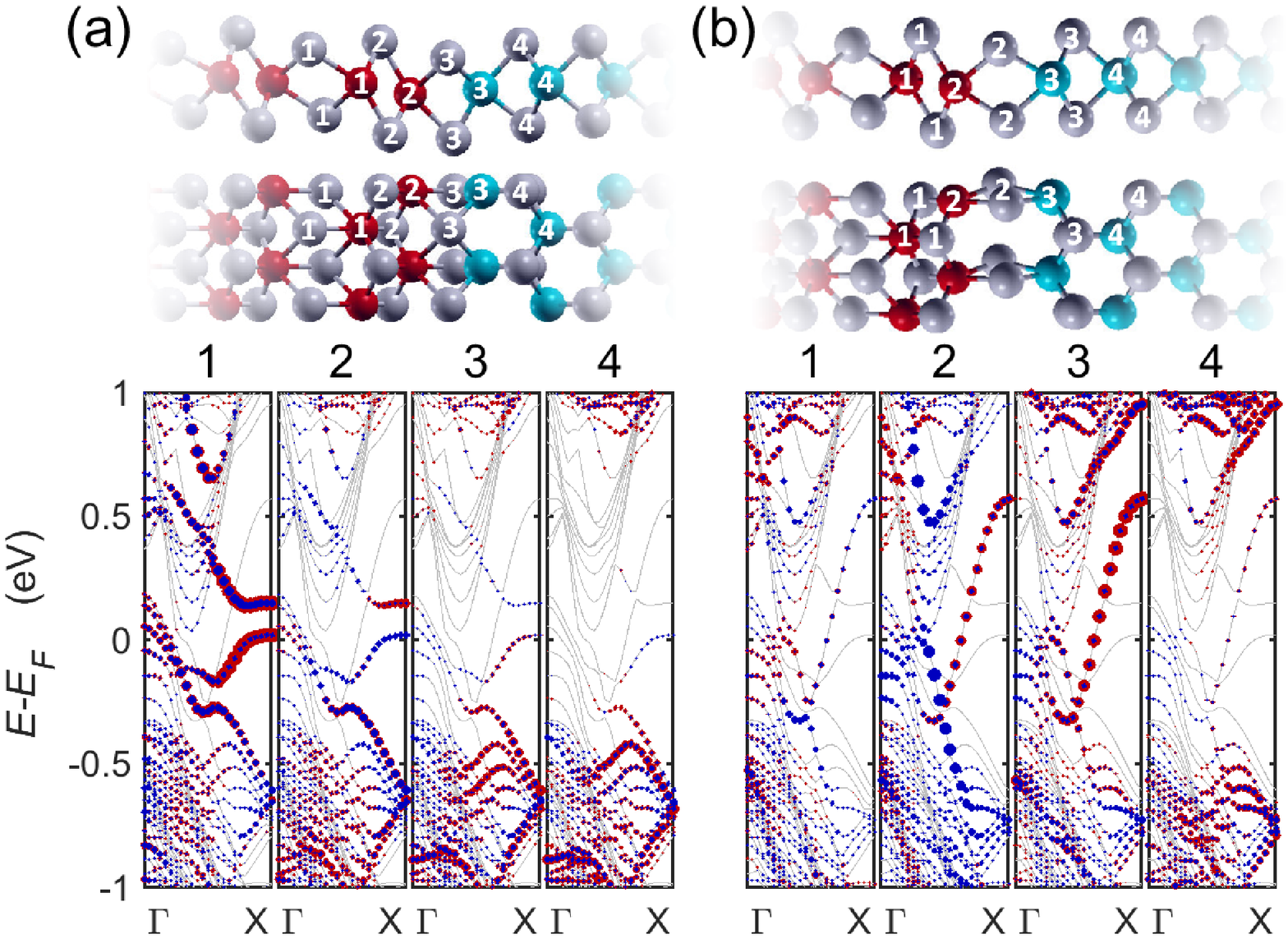}
\includegraphics[width= 3.5in]{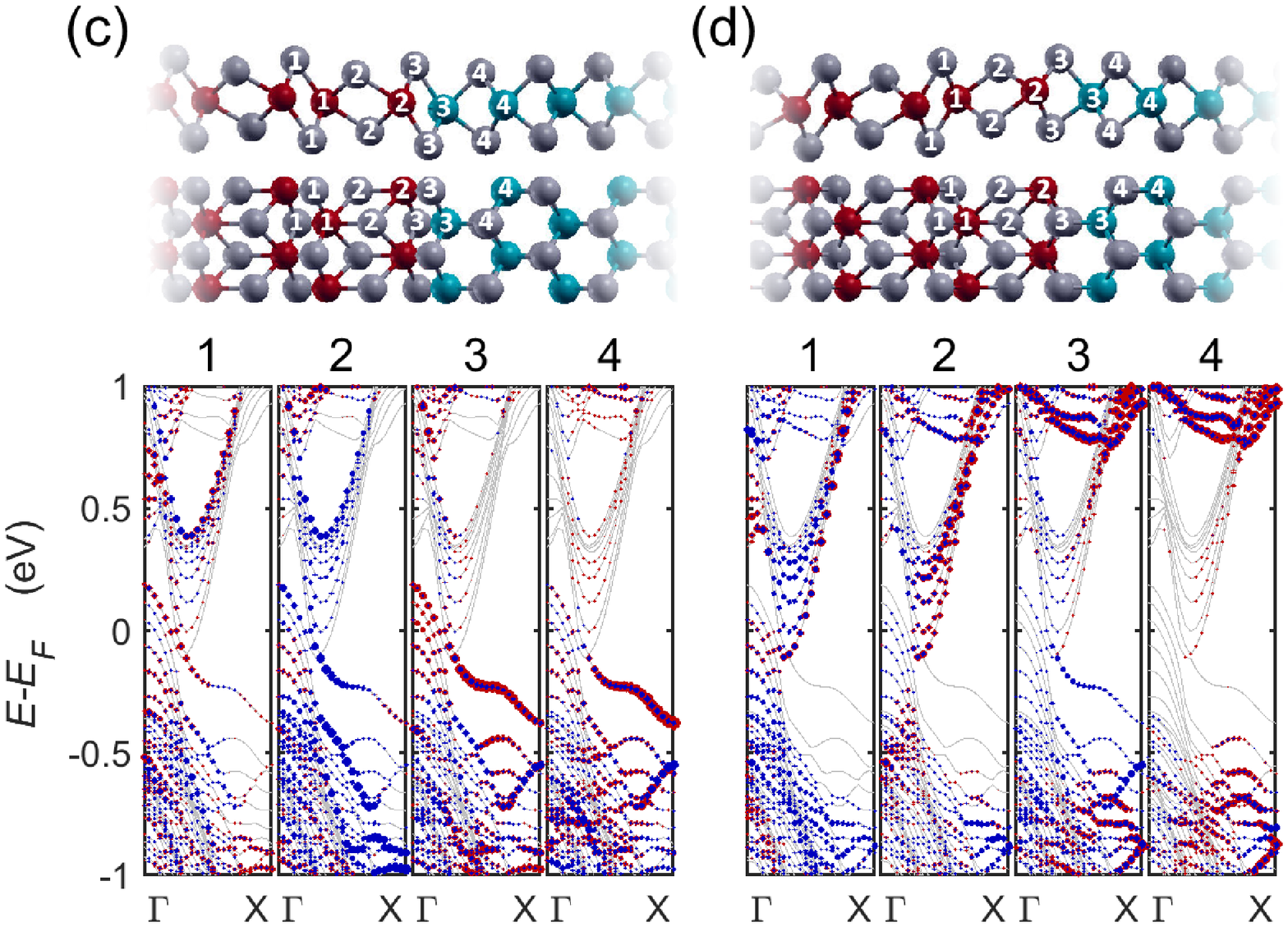}
\caption{Geometric and electronic structure of the interface between 1T'-WTe$_2$ and 2H-MoTe$_2$ for different edge terminations. Schematics of each junction include a side and top view and are displayed above the decomposition of Bloch states onto localized LOAs. Each sub-panel corresponds to atoms in different portions of the interface as labeled by numbers 1-4.
}\label{fig:interfaces}}
\end{figure}	

We characterize the electronic properties corresponding to different 1T'-WTe$_2$/2H-MoTe$_2$ edge terminations.  The geometric structure of the interface and the corresponding decompositions of Bloch states onto LAO are presented in Figure~\ref{fig:interfaces}. The $k_x$-dispersion of those states can change considerably depending on the interface details and appears to be related to the positions of the chalcogens between the 1T' and 2H TMDs. For instance, if the chalcogen atoms near the interface are on top of one another, as in the 2H structure, the interface state appears to be a single continuous state between the valence and conduction bands (see subplots 8-10 of Figure~\ref{fig:pdos} and Figure~\ref{fig:interfaces}b). If instead the chalcogens are not aligned, like in the 1T' structure, the interface states are spread out in multiple bands with some gaps, as in subplots 1-3 of Figure~\ref{fig:pdos} and Figure~\ref{fig:interfaces}a. We find that the original supercell is energetically favorable over the cases with unique edge terminations and computation of their average transmissions yield features similar to those observed in Figure~\ref{fig:2h_trans} (S7).   The presence and extension of MIGS at the interfaces, nonetheless, influence charge transport because they offer intermediate paths to conduction that effectively reduce the channel length and enhance tunneling.

\section{Conclusions}

In summary, we use first principles calculations to analyze the contact resistance of lateral heterojunctions formed by metallic 1T'-WTe$_2$ leads and various group VI TMD channels.   We find that systems with 1T'-MoTe$_2$ channels present almost ideal carrier injection, with a zero-bias contact resistance of approximately 5~$\Omega\cdot\mu$m. Degradation of the contact resistance for junctions with 1T' selenide channels is attributed to the weak overlap between modes in the electrode and channel. In turn, semiconducting 2H channels exhibit Schottky barriers for both electrons and holes in these lateral TMD heterostructures.  Furthermore, transmission near the valence or conduction bands yields contact resistances of about an order of magnitude lower than those found experimentally, signaling opportunities to improve contacts through phase-engineering and doping. The dominant  transport mechanism shifts from thermionic emission to tunneling for lengths shorter than about 3~nm.  We describe the dispersion of interface states in the metal-semiconducting systems and their influence in transport. Aside from the carrier injection, we show that the thermal stability of the semiconducting channel of MoTe$_2$, MoSe$_2$ and WSe$_2$ can be controlled by varying the channel length, enabling phase-change material platforms through the interplay of geometry and composition.

\section{Acknowledgements}

The authors would like to acknowledge the  Hopper HPC cluster for providing resources which have contributed to the research results reported within this paper. M.A.K. is thankful for startup funds at Auburn University.

\bibliography{mybib}

\begin{thebibliography}{65}%
\makeatletter
\providecommand \@ifxundefined [1]{%
 \@ifx{#1\undefined}
}%
\providecommand \@ifnum [1]{%
 \ifnum #1\expandafter \@firstoftwo
 \else \expandafter \@secondoftwo
 \fi
}%
\providecommand \@ifx [1]{%
 \ifx #1\expandafter \@firstoftwo
 \else \expandafter \@secondoftwo
 \fi
}%
\providecommand \natexlab [1]{#1}%
\providecommand \enquote  [1]{``#1''}%
\providecommand \bibnamefont  [1]{#1}%
\providecommand \bibfnamefont [1]{#1}%
\providecommand \citenamefont [1]{#1}%
\providecommand \href@noop [0]{\@secondoftwo}%
\providecommand \href [0]{\begingroup \@sanitize@url \@href}%
\providecommand \@href[1]{\@@startlink{#1}\@@href}%
\providecommand \@@href[1]{\endgroup#1\@@endlink}%
\providecommand \@sanitize@url [0]{\catcode `\\12\catcode `\$12\catcode
  `\&12\catcode `\#12\catcode `\^12\catcode `\_12\catcode `\%12\relax}%
\providecommand \@@startlink[1]{}%
\providecommand \@@endlink[0]{}%
\providecommand \url  [0]{\begingroup\@sanitize@url \@url }%
\providecommand \@url [1]{\endgroup\@href {#1}{\urlprefix }}%
\providecommand \urlprefix  [0]{URL }%
\providecommand \Eprint [0]{\href }%
\providecommand \doibase [0]{http://dx.doi.org/}%
\providecommand \selectlanguage [0]{\@gobble}%
\providecommand \bibinfo  [0]{\@secondoftwo}%
\providecommand \bibfield  [0]{\@secondoftwo}%
\providecommand \translation [1]{[#1]}%
\providecommand \BibitemOpen [0]{}%
\providecommand \bibitemStop [0]{}%
\providecommand \bibitemNoStop [0]{.\EOS\space}%
\providecommand \EOS [0]{\spacefactor3000\relax}%
\providecommand \BibitemShut  [1]{\csname bibitem#1\endcsname}%
\let\auto@bib@innerbib\@empty
\bibitem [{\citenamefont {Novoselov}\ \emph
  {et~al.}(2005{\natexlab{a}})\citenamefont {Novoselov}, \citenamefont {Jiang},
  \citenamefont {Schedin}, \citenamefont {Booth}, \citenamefont {Khotkevich},
  \citenamefont {Morozov},\ and\ \citenamefont {Geim}}]{novoselov2005a}%
  \BibitemOpen
  \bibfield  {author} {\bibinfo {author} {\bibfnamefont {K.~S.}\ \bibnamefont
  {Novoselov}}, \bibinfo {author} {\bibfnamefont {D.}~\bibnamefont {Jiang}},
  \bibinfo {author} {\bibfnamefont {F.}~\bibnamefont {Schedin}}, \bibinfo
  {author} {\bibfnamefont {T.~J.}\ \bibnamefont {Booth}}, \bibinfo {author}
  {\bibfnamefont {V.~V.}\ \bibnamefont {Khotkevich}}, \bibinfo {author}
  {\bibfnamefont {S.~V.}\ \bibnamefont {Morozov}}, \ and\ \bibinfo {author}
  {\bibfnamefont {A.~K.}\ \bibnamefont {Geim}},\ }\href@noop {} {\bibfield
  {journal} {\bibinfo  {journal} {Proc. Natl. Acad. Sci. USA}\ }\textbf
  {\bibinfo {volume} {102}},\ \bibinfo {pages} {10451} (\bibinfo {year}
  {2005}{\natexlab{a}})}\BibitemShut {NoStop}%
\bibitem [{\citenamefont {Gupta}\ \emph {et~al.}(2015)\citenamefont {Gupta},
  \citenamefont {Sakthivel},\ and\ \citenamefont {Seal}}]{gupta2015}%
  \BibitemOpen
  \bibfield  {author} {\bibinfo {author} {\bibfnamefont {A.}~\bibnamefont
  {Gupta}}, \bibinfo {author} {\bibfnamefont {T.}~\bibnamefont {Sakthivel}}, \
  and\ \bibinfo {author} {\bibfnamefont {S.}~\bibnamefont {Seal}},\ }\href
  {http://www.sciencedirect.com/science/article/pii/S0079642515000237}
  {\bibfield  {journal} {\bibinfo  {journal} {Progress in Materials Science}\
  }\textbf {\bibinfo {volume} {73}},\ \bibinfo {pages} {44 } (\bibinfo {year}
  {2015})}\BibitemShut {NoStop}%
\bibitem [{\citenamefont {Novoselov}\ \emph {et~al.}(2004)\citenamefont
  {Novoselov}, \citenamefont {Geim}, \citenamefont {Morozov}, \citenamefont
  {Jiang}, \citenamefont {Zhang}, \citenamefont {Dubonos}, \citenamefont
  {Grigorieva},\ and\ \citenamefont {Firsov}}]{Novoselov2004}%
  \BibitemOpen
  \bibfield  {author} {\bibinfo {author} {\bibfnamefont {K.~S.}\ \bibnamefont
  {Novoselov}}, \bibinfo {author} {\bibfnamefont {A.~K.}\ \bibnamefont {Geim}},
  \bibinfo {author} {\bibfnamefont {S.~V.}\ \bibnamefont {Morozov}}, \bibinfo
  {author} {\bibfnamefont {D.}~\bibnamefont {Jiang}}, \bibinfo {author}
  {\bibfnamefont {Y.}~\bibnamefont {Zhang}}, \bibinfo {author} {\bibfnamefont
  {S.~V.}\ \bibnamefont {Dubonos}}, \bibinfo {author} {\bibfnamefont {I.~V.}\
  \bibnamefont {Grigorieva}}, \ and\ \bibinfo {author} {\bibfnamefont {A.~A.}\
  \bibnamefont {Firsov}},\ }\href
  {http://science.sciencemag.org/content/306/5696/666} {\bibfield  {journal}
  {\bibinfo  {journal} {Science}\ }\textbf {\bibinfo {volume} {306}},\ \bibinfo
  {pages} {666} (\bibinfo {year} {2004})}\BibitemShut {NoStop}%
\bibitem [{\citenamefont {Novoselov}\ \emph
  {et~al.}(2005{\natexlab{b}})\citenamefont {Novoselov}, \citenamefont {Jiang},
  \citenamefont {Schedin}, \citenamefont {Booth}, \citenamefont {Khotkevich},
  \citenamefont {Morozov},\ and\ \citenamefont {Geim}}]{Novoselov2005}%
  \BibitemOpen
  \bibfield  {author} {\bibinfo {author} {\bibfnamefont {K.~S.}\ \bibnamefont
  {Novoselov}}, \bibinfo {author} {\bibfnamefont {D.}~\bibnamefont {Jiang}},
  \bibinfo {author} {\bibfnamefont {F.}~\bibnamefont {Schedin}}, \bibinfo
  {author} {\bibfnamefont {T.~J.}\ \bibnamefont {Booth}}, \bibinfo {author}
  {\bibfnamefont {V.~V.}\ \bibnamefont {Khotkevich}}, \bibinfo {author}
  {\bibfnamefont {S.~V.}\ \bibnamefont {Morozov}}, \ and\ \bibinfo {author}
  {\bibfnamefont {A.~K.}\ \bibnamefont {Geim}},\ }\href
  {https://www.pnas.org/content/102/30/10451} {\bibfield  {journal} {\bibinfo
  {journal} {Proceedings of the National Academy of Sciences}\ }\textbf
  {\bibinfo {volume} {102}},\ \bibinfo {pages} {10451} (\bibinfo {year}
  {2005}{\natexlab{b}})}\BibitemShut {NoStop}%
\bibitem [{\citenamefont {Radisavljevic}\ \emph {et~al.}(2011)\citenamefont
  {Radisavljevic}, \citenamefont {Radenovic}, \citenamefont {Brivio},
  \citenamefont {Giacometti},\ and\ \citenamefont {Kis}}]{Radisavljevic2011}%
  \BibitemOpen
  \bibfield  {author} {\bibinfo {author} {\bibfnamefont {B.}~\bibnamefont
  {Radisavljevic}}, \bibinfo {author} {\bibfnamefont {A.}~\bibnamefont
  {Radenovic}}, \bibinfo {author} {\bibfnamefont {J.}~\bibnamefont {Brivio}},
  \bibinfo {author} {\bibfnamefont {V.}~\bibnamefont {Giacometti}}, \ and\
  \bibinfo {author} {\bibfnamefont {A.}~\bibnamefont {Kis}},\ }\href@noop {}
  {\bibfield  {journal} {\bibinfo  {journal} {Nature nanotechnology}\ }\textbf
  {\bibinfo {volume} {6}},\ \bibinfo {pages} {147} (\bibinfo {year}
  {2011})}\BibitemShut {NoStop}%
\bibitem [{\citenamefont {Chen}\ \emph {et~al.}(2018)\citenamefont {Chen},
  \citenamefont {Li}, \citenamefont {Chen}, \citenamefont {Ong},\ and\
  \citenamefont {Zhao}}]{Chen2018}%
  \BibitemOpen
  \bibfield  {author} {\bibinfo {author} {\bibfnamefont {P.}~\bibnamefont
  {Chen}}, \bibinfo {author} {\bibfnamefont {N.}~\bibnamefont {Li}}, \bibinfo
  {author} {\bibfnamefont {X.}~\bibnamefont {Chen}}, \bibinfo {author}
  {\bibfnamefont {W.}~\bibnamefont {Ong}}, \ and\ \bibinfo {author}
  {\bibfnamefont {X.}~\bibnamefont {Zhao}},\ }\href@noop {} {\bibfield
  {journal} {\bibinfo  {journal} {2D Materials}\ }\textbf {\bibinfo {volume}
  {5}},\ \bibinfo {pages} {014002} (\bibinfo {year} {2018})}\BibitemShut
  {NoStop}%
\bibitem [{\citenamefont {Han}\ \emph {et~al.}(2008)\citenamefont {Han},
  \citenamefont {Wu}, \citenamefont {Zhu}, \citenamefont {Watanabe},\ and\
  \citenamefont {Taniguchi}}]{Han2008}%
  \BibitemOpen
  \bibfield  {author} {\bibinfo {author} {\bibfnamefont {W.-Q.}\ \bibnamefont
  {Han}}, \bibinfo {author} {\bibfnamefont {L.}~\bibnamefont {Wu}}, \bibinfo
  {author} {\bibfnamefont {Y.}~\bibnamefont {Zhu}}, \bibinfo {author}
  {\bibfnamefont {K.}~\bibnamefont {Watanabe}}, \ and\ \bibinfo {author}
  {\bibfnamefont {T.}~\bibnamefont {Taniguchi}},\ }\href
  {https://doi.org/10.1063/1.3041639} {\bibfield  {journal} {\bibinfo
  {journal} {Applied Physics Letters}\ }\textbf {\bibinfo {volume} {93}},\
  \bibinfo {pages} {223103} (\bibinfo {year} {2008})}\BibitemShut {NoStop}%
\bibitem [{\citenamefont {Hu}\ \emph {et~al.}(2012)\citenamefont {Hu},
  \citenamefont {Wen}, \citenamefont {Wang}, \citenamefont {Tan},\ and\
  \citenamefont {Xiao}}]{Hu2012}%
  \BibitemOpen
  \bibfield  {author} {\bibinfo {author} {\bibfnamefont {P.}~\bibnamefont
  {Hu}}, \bibinfo {author} {\bibfnamefont {Z.}~\bibnamefont {Wen}}, \bibinfo
  {author} {\bibfnamefont {L.}~\bibnamefont {Wang}}, \bibinfo {author}
  {\bibfnamefont {P.}~\bibnamefont {Tan}}, \ and\ \bibinfo {author}
  {\bibfnamefont {K.}~\bibnamefont {Xiao}},\ }\href
  {https://doi.org/10.1021/nn300889c} {\bibfield  {journal} {\bibinfo
  {journal} {ACS Nano}\ }\textbf {\bibinfo {volume} {6}},\ \bibinfo {pages}
  {5988} (\bibinfo {year} {2012})}\BibitemShut {NoStop}%
\bibitem [{\citenamefont {Chhowalla}\ \emph {et~al.}(2013)\citenamefont
  {Chhowalla}, \citenamefont {Shin}, \citenamefont {Eda}, \citenamefont {Li},
  \citenamefont {Loh},\ and\ \citenamefont {Zhang}}]{chhowalla2013}%
  \BibitemOpen
  \bibfield  {author} {\bibinfo {author} {\bibfnamefont {M.}~\bibnamefont
  {Chhowalla}}, \bibinfo {author} {\bibfnamefont {H.~S.}\ \bibnamefont {Shin}},
  \bibinfo {author} {\bibfnamefont {G.}~\bibnamefont {Eda}}, \bibinfo {author}
  {\bibfnamefont {L.-J.}\ \bibnamefont {Li}}, \bibinfo {author} {\bibfnamefont
  {K.~P.}\ \bibnamefont {Loh}}, \ and\ \bibinfo {author} {\bibfnamefont
  {H.}~\bibnamefont {Zhang}},\ }\href@noop {} {\bibfield  {journal} {\bibinfo
  {journal} {Nature Chem.}\ }\textbf {\bibinfo {volume} {5}},\ \bibinfo {pages}
  {263} (\bibinfo {year} {2013})}\BibitemShut {NoStop}%
\bibitem [{\citenamefont {Wang}\ \emph {et~al.}(2012)\citenamefont {Wang},
  \citenamefont {Kalantar-Zadeh}, \citenamefont {Kis}, \citenamefont
  {Coleman},\ and\ \citenamefont {Strano}}]{wang2012}%
  \BibitemOpen
  \bibfield  {author} {\bibinfo {author} {\bibfnamefont {Q.~H.}\ \bibnamefont
  {Wang}}, \bibinfo {author} {\bibfnamefont {K.}~\bibnamefont
  {Kalantar-Zadeh}}, \bibinfo {author} {\bibfnamefont {A.}~\bibnamefont {Kis}},
  \bibinfo {author} {\bibfnamefont {J.~N.}\ \bibnamefont {Coleman}}, \ and\
  \bibinfo {author} {\bibfnamefont {M.~S.}\ \bibnamefont {Strano}},\
  }\href@noop {} {\bibfield  {journal} {\bibinfo  {journal} {Nature Nanotech.}\
  }\textbf {\bibinfo {volume} {7}},\ \bibinfo {pages} {699} (\bibinfo {year}
  {2012})}\BibitemShut {NoStop}%
\bibitem [{\citenamefont {Cao}\ \emph {et~al.}(2012)\citenamefont {Cao},
  \citenamefont {Wang}, \citenamefont {Han}, \citenamefont {Ye}, \citenamefont
  {Zhu}, \citenamefont {Shi}, \citenamefont {Niu}, \citenamefont {Tan},
  \citenamefont {Wang}, \citenamefont {Liu},\ and\ \citenamefont
  {Feng}}]{cao2012}%
  \BibitemOpen
  \bibfield  {author} {\bibinfo {author} {\bibfnamefont {T.}~\bibnamefont
  {Cao}}, \bibinfo {author} {\bibfnamefont {G.}~\bibnamefont {Wang}}, \bibinfo
  {author} {\bibfnamefont {W.}~\bibnamefont {Han}}, \bibinfo {author}
  {\bibfnamefont {H.}~\bibnamefont {Ye}}, \bibinfo {author} {\bibfnamefont
  {C.}~\bibnamefont {Zhu}}, \bibinfo {author} {\bibfnamefont {J.}~\bibnamefont
  {Shi}}, \bibinfo {author} {\bibfnamefont {Q.}~\bibnamefont {Niu}}, \bibinfo
  {author} {\bibfnamefont {P.}~\bibnamefont {Tan}}, \bibinfo {author}
  {\bibfnamefont {E.}~\bibnamefont {Wang}}, \bibinfo {author} {\bibfnamefont
  {B.}~\bibnamefont {Liu}}, \ and\ \bibinfo {author} {\bibfnamefont
  {J.}~\bibnamefont {Feng}},\ }\href@noop {} {\bibfield  {journal} {\bibinfo
  {journal} {Nature Communications}\ }\textbf {\bibinfo {volume} {3}},\
  \bibinfo {pages} {887} (\bibinfo {year} {2012})}\BibitemShut {NoStop}%
\bibitem [{\citenamefont {Mak}\ \emph {et~al.}(2012)\citenamefont {Mak},
  \citenamefont {He}, \citenamefont {Shan},\ and\ \citenamefont
  {Heinz}}]{mak2012}%
  \BibitemOpen
  \bibfield  {author} {\bibinfo {author} {\bibfnamefont {K.~F.}\ \bibnamefont
  {Mak}}, \bibinfo {author} {\bibfnamefont {K.}~\bibnamefont {He}}, \bibinfo
  {author} {\bibfnamefont {J.}~\bibnamefont {Shan}}, \ and\ \bibinfo {author}
  {\bibfnamefont {T.~F.}\ \bibnamefont {Heinz}},\ }\href@noop {} {\bibfield
  {journal} {\bibinfo  {journal} {Nat Nano}\ }\textbf {\bibinfo {volume} {7}},\
  \bibinfo {pages} {494} (\bibinfo {year} {2012})}\BibitemShut {NoStop}%
\bibitem [{\citenamefont {Duerloo}\ \emph {et~al.}(2013)\citenamefont
  {Duerloo}, \citenamefont {Li},\ and\ \citenamefont {Reed}}]{duerloo2013}%
  \BibitemOpen
  \bibfield  {author} {\bibinfo {author} {\bibfnamefont {K.-A.}\ \bibnamefont
  {Duerloo}}, \bibinfo {author} {\bibfnamefont {Y.}~\bibnamefont {Li}}, \ and\
  \bibinfo {author} {\bibfnamefont {E.}~\bibnamefont {Reed}},\ }\href@noop {}
  {\bibfield  {journal} {\bibinfo  {journal} {Nature Communications}\ }\textbf
  {\bibinfo {volume} {5}},\ \bibinfo {pages} {4214} (\bibinfo {year}
  {2013})}\BibitemShut {NoStop}%
\bibitem [{\citenamefont {Zeng}\ and\ \citenamefont
  {Cui}(2015)}]{ChemReview2015}%
  \BibitemOpen
  \bibfield  {author} {\bibinfo {author} {\bibfnamefont {H.}~\bibnamefont
  {Zeng}}\ and\ \bibinfo {author} {\bibfnamefont {X.}~\bibnamefont {Cui}},\
  }\href {\doibase 10.1039/C4CS00265B} {\bibfield  {journal} {\bibinfo
  {journal} {Chem. Soc. Rev.}\ }\textbf {\bibinfo {volume} {44}},\ \bibinfo
  {pages} {2629} (\bibinfo {year} {2015})}\BibitemShut {NoStop}%
\bibitem [{\citenamefont {Duerloo}\ and\ \citenamefont
  {Reed}(2016)}]{duerloo2016}%
  \BibitemOpen
  \bibfield  {author} {\bibinfo {author} {\bibfnamefont {K.-A.~N.}\
  \bibnamefont {Duerloo}}\ and\ \bibinfo {author} {\bibfnamefont {E.~J.}\
  \bibnamefont {Reed}},\ }\href {https://doi.org/10.1021/acsnano.5b04359}
  {\bibfield  {journal} {\bibinfo  {journal} {ACS Nano}\ }\textbf {\bibinfo
  {volume} {10}},\ \bibinfo {pages} {289} (\bibinfo {year} {2016})}\BibitemShut
  {NoStop}%
\bibitem [{\citenamefont {Sebastian}\ \emph {et~al.}(2018)\citenamefont
  {Sebastian}, \citenamefont {Le~Gallo}, \citenamefont {Burr}, \citenamefont
  {Kim}, \citenamefont {BrightSky},\ and\ \citenamefont
  {Eleftheriou}}]{sebastian2018}%
  \BibitemOpen
  \bibfield  {author} {\bibinfo {author} {\bibfnamefont {A.}~\bibnamefont
  {Sebastian}}, \bibinfo {author} {\bibfnamefont {M.}~\bibnamefont {Le~Gallo}},
  \bibinfo {author} {\bibfnamefont {G.~W.}\ \bibnamefont {Burr}}, \bibinfo
  {author} {\bibfnamefont {S.}~\bibnamefont {Kim}}, \bibinfo {author}
  {\bibfnamefont {M.}~\bibnamefont {BrightSky}}, \ and\ \bibinfo {author}
  {\bibfnamefont {E.}~\bibnamefont {Eleftheriou}},\ }\href@noop {} {\bibfield
  {journal} {\bibinfo  {journal} {Journal of Applied Physics}\ }\textbf
  {\bibinfo {volume} {124}},\ \bibinfo {pages} {111101} (\bibinfo {year}
  {2018})}\BibitemShut {NoStop}%
\bibitem [{\citenamefont {Raoux}\ \emph {et~al.}(2014)\citenamefont {Raoux},
  \citenamefont {Xiong}, \citenamefont {Wuttig},\ and\ \citenamefont
  {Pop}}]{raoux2014}%
  \BibitemOpen
  \bibfield  {author} {\bibinfo {author} {\bibfnamefont {S.}~\bibnamefont
  {Raoux}}, \bibinfo {author} {\bibfnamefont {F.}~\bibnamefont {Xiong}},
  \bibinfo {author} {\bibfnamefont {M.}~\bibnamefont {Wuttig}}, \ and\ \bibinfo
  {author} {\bibfnamefont {E.}~\bibnamefont {Pop}},\ }\href@noop {} {\bibfield
  {journal} {\bibinfo  {journal} {MRS Bulletin}\ }\textbf {\bibinfo {volume}
  {39}},\ \bibinfo {pages} {703} (\bibinfo {year} {2014})}\BibitemShut
  {NoStop}%
\bibitem [{\citenamefont {Yoon}\ \emph {et~al.}(2011)\citenamefont {Yoon},
  \citenamefont {Ganapathi},\ and\ \citenamefont {Salahuddin}}]{yoon2011}%
  \BibitemOpen
  \bibfield  {author} {\bibinfo {author} {\bibfnamefont {Y.}~\bibnamefont
  {Yoon}}, \bibinfo {author} {\bibfnamefont {K.}~\bibnamefont {Ganapathi}}, \
  and\ \bibinfo {author} {\bibfnamefont {S.}~\bibnamefont {Salahuddin}},\
  }\href {https://doi.org/10.1021/nl2018178} {\bibfield  {journal} {\bibinfo
  {journal} {Nano Letters}\ }\textbf {\bibinfo {volume} {11}},\ \bibinfo
  {pages} {3768} (\bibinfo {year} {2011})}\BibitemShut {NoStop}%
\bibitem [{\citenamefont {Allain}\ and\ \citenamefont
  {Kis}(2014)}]{Allain2014}%
  \BibitemOpen
  \bibfield  {author} {\bibinfo {author} {\bibfnamefont {A.}~\bibnamefont
  {Allain}}\ and\ \bibinfo {author} {\bibfnamefont {A.}~\bibnamefont {Kis}},\
  }\href {https://doi.org/10.1021/nn5021538} {\bibfield  {journal} {\bibinfo
  {journal} {ACS Nano}\ }\textbf {\bibinfo {volume} {8}},\ \bibinfo {pages}
  {7180} (\bibinfo {year} {2014})}\BibitemShut {NoStop}%
\bibitem [{\citenamefont {L{\'e}onard}\ and\ \citenamefont
  {Talin}(2011)}]{leonard2011}%
  \BibitemOpen
  \bibfield  {author} {\bibinfo {author} {\bibfnamefont {F.}~\bibnamefont
  {L{\'e}onard}}\ and\ \bibinfo {author} {\bibfnamefont {A.~A.}\ \bibnamefont
  {Talin}},\ }\href {https://doi.org/10.1038/nnano.2011.196} {\bibfield
  {journal} {\bibinfo  {journal} {Nature Nanotechnology}\ }\textbf {\bibinfo
  {volume} {6}},\ \bibinfo {pages} {773} (\bibinfo {year} {2011})}\BibitemShut
  {NoStop}%
\bibitem [{\citenamefont {Das}\ \emph {et~al.}(2013)\citenamefont {Das},
  \citenamefont {Chen}, \citenamefont {Penumatcha},\ and\ \citenamefont
  {Appenzeller}}]{das2013a}%
  \BibitemOpen
  \bibfield  {author} {\bibinfo {author} {\bibfnamefont {S.}~\bibnamefont
  {Das}}, \bibinfo {author} {\bibfnamefont {H.-Y.}\ \bibnamefont {Chen}},
  \bibinfo {author} {\bibfnamefont {A.~V.}\ \bibnamefont {Penumatcha}}, \ and\
  \bibinfo {author} {\bibfnamefont {J.}~\bibnamefont {Appenzeller}},\
  }\bibfield  {booktitle} {\emph {\bibinfo {booktitle} {Nano Letters}},\ }\href
  {https://doi.org/10.1021/nl303583v} {\bibfield  {journal} {\bibinfo
  {journal} {Nano Letters}\ }\textbf {\bibinfo {volume} {13}},\ \bibinfo
  {pages} {100} (\bibinfo {year} {2013})}\BibitemShut {NoStop}%
\bibitem [{\citenamefont {Chuang}\ \emph {et~al.}(2014)\citenamefont {Chuang},
  \citenamefont {Battaglia}, \citenamefont {Azcatl}, \citenamefont {McDonnell},
  \citenamefont {Kang}, \citenamefont {Yin}, \citenamefont {Tosun},
  \citenamefont {Kapadia}, \citenamefont {Fang}, \citenamefont {Wallace},\ and\
  \citenamefont {Javey}}]{chuang2014}%
  \BibitemOpen
  \bibfield  {author} {\bibinfo {author} {\bibfnamefont {S.}~\bibnamefont
  {Chuang}}, \bibinfo {author} {\bibfnamefont {C.}~\bibnamefont {Battaglia}},
  \bibinfo {author} {\bibfnamefont {A.}~\bibnamefont {Azcatl}}, \bibinfo
  {author} {\bibfnamefont {S.}~\bibnamefont {McDonnell}}, \bibinfo {author}
  {\bibfnamefont {J.~S.}\ \bibnamefont {Kang}}, \bibinfo {author}
  {\bibfnamefont {X.}~\bibnamefont {Yin}}, \bibinfo {author} {\bibfnamefont
  {M.}~\bibnamefont {Tosun}}, \bibinfo {author} {\bibfnamefont
  {R.}~\bibnamefont {Kapadia}}, \bibinfo {author} {\bibfnamefont
  {H.}~\bibnamefont {Fang}}, \bibinfo {author} {\bibfnamefont {R.~M.}\
  \bibnamefont {Wallace}}, \ and\ \bibinfo {author} {\bibfnamefont
  {A.}~\bibnamefont {Javey}},\ }\href@noop {} {\bibfield  {journal} {\bibinfo
  {journal} {Nano Lett.}\ }\textbf {\bibinfo {volume} {14}},\ \bibinfo {pages}
  {1337} (\bibinfo {year} {2014})}\BibitemShut {NoStop}%
\bibitem [{\citenamefont {Kumar}\ \emph {et~al.}(2015)\citenamefont {Kumar},
  \citenamefont {Kuroda}, \citenamefont {Bellus}, \citenamefont {Han},\ and\
  \citenamefont {Chiu}}]{kumar2015}%
  \BibitemOpen
  \bibfield  {author} {\bibinfo {author} {\bibfnamefont {J.}~\bibnamefont
  {Kumar}}, \bibinfo {author} {\bibfnamefont {M.~A.}\ \bibnamefont {Kuroda}},
  \bibinfo {author} {\bibfnamefont {M.~Z.}\ \bibnamefont {Bellus}}, \bibinfo
  {author} {\bibfnamefont {S.-J.}\ \bibnamefont {Han}}, \ and\ \bibinfo
  {author} {\bibfnamefont {H.-Y.}\ \bibnamefont {Chiu}},\ }\href@noop {}
  {\bibfield  {journal} {\bibinfo  {journal} {Appl. Phys. Lett.}\ }\textbf
  {\bibinfo {volume} {106}},\ \bibinfo {eid} {123508} (\bibinfo {year}
  {2015})}\BibitemShut {NoStop}%
\bibitem [{\citenamefont {Kim}\ \emph {et~al.}(2017)\citenamefont {Kim},
  \citenamefont {Moon}, \citenamefont {Lee}, \citenamefont {Choi},
  \citenamefont {Ahmed}, \citenamefont {Nam}, \citenamefont {Cho},
  \citenamefont {Shin}, \citenamefont {Park},\ and\ \citenamefont
  {Yoo}}]{Kim2017}%
  \BibitemOpen
  \bibfield  {author} {\bibinfo {author} {\bibfnamefont {C.}~\bibnamefont
  {Kim}}, \bibinfo {author} {\bibfnamefont {I.}~\bibnamefont {Moon}}, \bibinfo
  {author} {\bibfnamefont {D.}~\bibnamefont {Lee}}, \bibinfo {author}
  {\bibfnamefont {M.~S.}\ \bibnamefont {Choi}}, \bibinfo {author}
  {\bibfnamefont {F.}~\bibnamefont {Ahmed}}, \bibinfo {author} {\bibfnamefont
  {S.}~\bibnamefont {Nam}}, \bibinfo {author} {\bibfnamefont {Y.}~\bibnamefont
  {Cho}}, \bibinfo {author} {\bibfnamefont {H.-J.}\ \bibnamefont {Shin}},
  \bibinfo {author} {\bibfnamefont {S.}~\bibnamefont {Park}}, \ and\ \bibinfo
  {author} {\bibfnamefont {W.~J.}\ \bibnamefont {Yoo}},\ }\href
  {https://doi.org/10.1021/acsnano.6b07159} {\bibfield  {journal} {\bibinfo
  {journal} {ACS Nano}\ }\textbf {\bibinfo {volume} {11}},\ \bibinfo {pages}
  {1588} (\bibinfo {year} {2017})}\BibitemShut {NoStop}%
\bibitem [{\citenamefont {Mahjouri-Samani}\ \emph {et~al.}(2015)\citenamefont
  {Mahjouri-Samani}, \citenamefont {Lin}, \citenamefont {Wang}, \citenamefont
  {Lupini}, \citenamefont {Lee}, \citenamefont {Basile}, \citenamefont
  {Boulesbaa}, \citenamefont {Rouleau}, \citenamefont {Puretzky}, \citenamefont
  {Ivanov}, \citenamefont {Xiao}, \citenamefont {Yoon},\ and\ \citenamefont
  {Geohegan}}]{mahjouri2015}%
  \BibitemOpen
  \bibfield  {author} {\bibinfo {author} {\bibfnamefont {M.}~\bibnamefont
  {Mahjouri-Samani}}, \bibinfo {author} {\bibfnamefont {M.-W.}\ \bibnamefont
  {Lin}}, \bibinfo {author} {\bibfnamefont {K.}~\bibnamefont {Wang}}, \bibinfo
  {author} {\bibfnamefont {A.~R.}\ \bibnamefont {Lupini}}, \bibinfo {author}
  {\bibfnamefont {J.}~\bibnamefont {Lee}}, \bibinfo {author} {\bibfnamefont
  {L.}~\bibnamefont {Basile}}, \bibinfo {author} {\bibfnamefont
  {A.}~\bibnamefont {Boulesbaa}}, \bibinfo {author} {\bibfnamefont {C.~M.}\
  \bibnamefont {Rouleau}}, \bibinfo {author} {\bibfnamefont {A.~A.}\
  \bibnamefont {Puretzky}}, \bibinfo {author} {\bibfnamefont {I.~N.}\
  \bibnamefont {Ivanov}}, \bibinfo {author} {\bibfnamefont {K.}~\bibnamefont
  {Xiao}}, \bibinfo {author} {\bibfnamefont {M.}~\bibnamefont {Yoon}}, \ and\
  \bibinfo {author} {\bibfnamefont {D.~B.}\ \bibnamefont {Geohegan}},\
  }\href@noop {} {\bibfield  {journal} {\bibinfo  {journal} {Nature
  Communications}\ }\textbf {\bibinfo {volume} {6}},\ \bibinfo {pages} {7749}
  (\bibinfo {year} {2015})}\BibitemShut {NoStop}%
\bibitem [{\citenamefont {Li}\ \emph {et~al.}(2015)\citenamefont {Li},
  \citenamefont {Shi}, \citenamefont {Cheng}, \citenamefont {Lu}, \citenamefont
  {Lin}, \citenamefont {Tang}, \citenamefont {Tsai}, \citenamefont {Chu},
  \citenamefont {Wei}, \citenamefont {He}, \citenamefont {Chang}, \citenamefont
  {Suenaga},\ and\ \citenamefont {Li}}]{li2015a}%
  \BibitemOpen
  \bibfield  {author} {\bibinfo {author} {\bibfnamefont {M.-Y.}\ \bibnamefont
  {Li}}, \bibinfo {author} {\bibfnamefont {Y.}~\bibnamefont {Shi}}, \bibinfo
  {author} {\bibfnamefont {C.-C.}\ \bibnamefont {Cheng}}, \bibinfo {author}
  {\bibfnamefont {L.-S.}\ \bibnamefont {Lu}}, \bibinfo {author} {\bibfnamefont
  {Y.-C.}\ \bibnamefont {Lin}}, \bibinfo {author} {\bibfnamefont {H.-L.}\
  \bibnamefont {Tang}}, \bibinfo {author} {\bibfnamefont {M.-L.}\ \bibnamefont
  {Tsai}}, \bibinfo {author} {\bibfnamefont {C.-W.}\ \bibnamefont {Chu}},
  \bibinfo {author} {\bibfnamefont {K.-H.}\ \bibnamefont {Wei}}, \bibinfo
  {author} {\bibfnamefont {J.-H.}\ \bibnamefont {He}}, \bibinfo {author}
  {\bibfnamefont {W.-H.}\ \bibnamefont {Chang}}, \bibinfo {author}
  {\bibfnamefont {K.}~\bibnamefont {Suenaga}}, \ and\ \bibinfo {author}
  {\bibfnamefont {L.-J.}\ \bibnamefont {Li}},\ }\href
  {http://science.sciencemag.org/content/349/6247/524} {\bibfield  {journal}
  {\bibinfo  {journal} {Science}\ }\textbf {\bibinfo {volume} {349}},\ \bibinfo
  {pages} {524} (\bibinfo {year} {2015})}\BibitemShut {NoStop}%
\bibitem [{\citenamefont {Cho}\ \emph {et~al.}(2015)\citenamefont {Cho},
  \citenamefont {Kim}, \citenamefont {Kim}, \citenamefont {Zhao}, \citenamefont
  {Seok}, \citenamefont {Keum}, \citenamefont {Baik}, \citenamefont {Choe},
  \citenamefont {Chang}, \citenamefont {Suenaga}, \citenamefont {Kim},
  \citenamefont {Lee},\ and\ \citenamefont {Yang}}]{Cho2015}%
  \BibitemOpen
  \bibfield  {author} {\bibinfo {author} {\bibfnamefont {S.}~\bibnamefont
  {Cho}}, \bibinfo {author} {\bibfnamefont {S.}~\bibnamefont {Kim}}, \bibinfo
  {author} {\bibfnamefont {J.~H.}\ \bibnamefont {Kim}}, \bibinfo {author}
  {\bibfnamefont {J.}~\bibnamefont {Zhao}}, \bibinfo {author} {\bibfnamefont
  {J.}~\bibnamefont {Seok}}, \bibinfo {author} {\bibfnamefont {D.~H.}\
  \bibnamefont {Keum}}, \bibinfo {author} {\bibfnamefont {J.}~\bibnamefont
  {Baik}}, \bibinfo {author} {\bibfnamefont {D.-H.}\ \bibnamefont {Choe}},
  \bibinfo {author} {\bibfnamefont {K.~J.}\ \bibnamefont {Chang}}, \bibinfo
  {author} {\bibfnamefont {K.}~\bibnamefont {Suenaga}}, \bibinfo {author}
  {\bibfnamefont {S.~W.}\ \bibnamefont {Kim}}, \bibinfo {author} {\bibfnamefont
  {Y.~H.}\ \bibnamefont {Lee}}, \ and\ \bibinfo {author} {\bibfnamefont
  {H.}~\bibnamefont {Yang}},\ }\href
  {http://science.sciencemag.org/content/349/6248/625} {\bibfield  {journal}
  {\bibinfo  {journal} {Science}\ }\textbf {\bibinfo {volume} {349}},\ \bibinfo
  {pages} {625} (\bibinfo {year} {2015})}\BibitemShut {NoStop}%
\bibitem [{\citenamefont {Chen}\ \emph {et~al.}(2013)\citenamefont {Chen},
  \citenamefont {Odenthal}, \citenamefont {Swartz}, \citenamefont {Floyd},
  \citenamefont {Wen}, \citenamefont {Luo},\ and\ \citenamefont
  {Kawakami}}]{Chen2013}%
  \BibitemOpen
  \bibfield  {author} {\bibinfo {author} {\bibfnamefont {J.-R.}\ \bibnamefont
  {Chen}}, \bibinfo {author} {\bibfnamefont {P.~M.}\ \bibnamefont {Odenthal}},
  \bibinfo {author} {\bibfnamefont {A.~G.}\ \bibnamefont {Swartz}}, \bibinfo
  {author} {\bibfnamefont {G.~C.}\ \bibnamefont {Floyd}}, \bibinfo {author}
  {\bibfnamefont {H.}~\bibnamefont {Wen}}, \bibinfo {author} {\bibfnamefont
  {K.~Y.}\ \bibnamefont {Luo}}, \ and\ \bibinfo {author} {\bibfnamefont
  {R.~K.}\ \bibnamefont {Kawakami}},\ }\href
  {https://doi.org/10.1021/nl4010157} {\bibfield  {journal} {\bibinfo
  {journal} {Nano Letters}\ }\textbf {\bibinfo {volume} {13}},\ \bibinfo
  {pages} {3106} (\bibinfo {year} {2013})}\BibitemShut {NoStop}%
\bibitem [{\citenamefont {Popov}\ \emph {et~al.}(2012)\citenamefont {Popov},
  \citenamefont {Seifert},\ and\ \citenamefont {Tomanek}}]{Popov2012}%
  \BibitemOpen
  \bibfield  {author} {\bibinfo {author} {\bibfnamefont {I.}~\bibnamefont
  {Popov}}, \bibinfo {author} {\bibfnamefont {G.}~\bibnamefont {Seifert}}, \
  and\ \bibinfo {author} {\bibfnamefont {D.}~\bibnamefont {Tomanek}},\
  }\href@noop {} {\bibfield  {journal} {\bibinfo  {journal} {Physical Review
  Letters}\ }\textbf {\bibinfo {volume} {108}},\ \bibinfo {pages} {156802}
  (\bibinfo {year} {2012})}\BibitemShut {NoStop}%
\bibitem [{\citenamefont {Kappera}\ \emph {et~al.}(2014)\citenamefont
  {Kappera}, \citenamefont {Voiry}, \citenamefont {Yalcin}, \citenamefont
  {Branch}, \citenamefont {Gupta}, \citenamefont {Mohite},\ and\ \citenamefont
  {Chhowalla}}]{kappera2014}%
  \BibitemOpen
  \bibfield  {author} {\bibinfo {author} {\bibfnamefont {R.}~\bibnamefont
  {Kappera}}, \bibinfo {author} {\bibfnamefont {D.}~\bibnamefont {Voiry}},
  \bibinfo {author} {\bibfnamefont {S.~E.}\ \bibnamefont {Yalcin}}, \bibinfo
  {author} {\bibfnamefont {B.}~\bibnamefont {Branch}}, \bibinfo {author}
  {\bibfnamefont {G.}~\bibnamefont {Gupta}}, \bibinfo {author} {\bibfnamefont
  {A.~D.}\ \bibnamefont {Mohite}}, \ and\ \bibinfo {author} {\bibfnamefont
  {M.}~\bibnamefont {Chhowalla}},\ }\href@noop {} {\bibfield  {journal}
  {\bibinfo  {journal} {Nat Mater}\ }\textbf {\bibinfo {volume} {13}},\
  \bibinfo {pages} {1128} (\bibinfo {year} {2014})}\BibitemShut {NoStop}%
\bibitem [{\citenamefont {Duran~Retamal}\ \emph {et~al.}(2018)\citenamefont
  {Duran~Retamal}, \citenamefont {Periyanagounder}, \citenamefont {Ke},
  \citenamefont {Tsai},\ and\ \citenamefont {He}}]{Retamal2018}%
  \BibitemOpen
  \bibfield  {author} {\bibinfo {author} {\bibfnamefont {J.~R.}\ \bibnamefont
  {Duran~Retamal}}, \bibinfo {author} {\bibfnamefont {D.}~\bibnamefont
  {Periyanagounder}}, \bibinfo {author} {\bibfnamefont {J.-J.}\ \bibnamefont
  {Ke}}, \bibinfo {author} {\bibfnamefont {M.-L.}\ \bibnamefont {Tsai}}, \ and\
  \bibinfo {author} {\bibfnamefont {J.-H.}\ \bibnamefont {He}},\ }\href
  {\doibase 10.1039/C8SC02609B} {\bibfield  {journal} {\bibinfo  {journal}
  {Chem. Sci.}\ }\textbf {\bibinfo {volume} {9}},\ \bibinfo {pages} {7727}
  (\bibinfo {year} {2018})}\BibitemShut {NoStop}%
\bibitem [{\citenamefont {Sung}\ \emph {et~al.}(2017)\citenamefont {Sung},
  \citenamefont {Heo}, \citenamefont {Si}, \citenamefont {Kim}, \citenamefont
  {Noh}, \citenamefont {Song}, \citenamefont {Kim}, \citenamefont {Lee},
  \citenamefont {Seo}, \citenamefont {Kim}, \citenamefont {Kim}, \citenamefont
  {Yeom}, \citenamefont {Kim}, \citenamefont {Choi}, \citenamefont {Kim},\ and\
  \citenamefont {Jo}}]{Sung2017}%
  \BibitemOpen
  \bibfield  {author} {\bibinfo {author} {\bibfnamefont {J.}~\bibnamefont
  {Sung}}, \bibinfo {author} {\bibfnamefont {H.}~\bibnamefont {Heo}}, \bibinfo
  {author} {\bibfnamefont {S.}~\bibnamefont {Si}}, \bibinfo {author}
  {\bibfnamefont {Y.}~\bibnamefont {Kim}}, \bibinfo {author} {\bibfnamefont
  {H.}~\bibnamefont {Noh}}, \bibinfo {author} {\bibfnamefont {K.}~\bibnamefont
  {Song}}, \bibinfo {author} {\bibfnamefont {J.}~\bibnamefont {Kim}}, \bibinfo
  {author} {\bibfnamefont {C.-S.}\ \bibnamefont {Lee}}, \bibinfo {author}
  {\bibfnamefont {S.-Y.}\ \bibnamefont {Seo}}, \bibinfo {author} {\bibfnamefont
  {D.-H.}\ \bibnamefont {Kim}}, \bibinfo {author} {\bibfnamefont
  {H.}~\bibnamefont {Kim}}, \bibinfo {author} {\bibfnamefont {H.}~\bibnamefont
  {Yeom}}, \bibinfo {author} {\bibfnamefont {T.-H.}\ \bibnamefont {Kim}},
  \bibinfo {author} {\bibfnamefont {S.-Y.}\ \bibnamefont {Choi}}, \bibinfo
  {author} {\bibfnamefont {J.-S.}\ \bibnamefont {Kim}}, \ and\ \bibinfo
  {author} {\bibfnamefont {M.-H.}\ \bibnamefont {Jo}},\ }\href@noop {}
  {\bibfield  {journal} {\bibinfo  {journal} {Nature Nanotechnology}\ }\textbf
  {\bibinfo {volume} {12}} (\bibinfo {year} {2017})}\BibitemShut {NoStop}%
\bibitem [{\citenamefont {Yang}\ \emph {et~al.}(2017)\citenamefont {Yang},
  \citenamefont {Zhang}, \citenamefont {Li}, \citenamefont {Cheng},
  \citenamefont {Xie},\ and\ \citenamefont {Chang}}]{Yang2017}%
  \BibitemOpen
  \bibfield  {author} {\bibinfo {author} {\bibfnamefont {L.}~\bibnamefont
  {Yang}}, \bibinfo {author} {\bibfnamefont {W.}~\bibnamefont {Zhang}},
  \bibinfo {author} {\bibfnamefont {J.}~\bibnamefont {Li}}, \bibinfo {author}
  {\bibfnamefont {S.}~\bibnamefont {Cheng}}, \bibinfo {author} {\bibfnamefont
  {Z.}~\bibnamefont {Xie}}, \ and\ \bibinfo {author} {\bibfnamefont
  {H.}~\bibnamefont {Chang}},\ }\href {https://doi.org/10.1021/acsnano.6b08109}
  {\bibfield  {journal} {\bibinfo  {journal} {ACS Nano}\ }\textbf {\bibinfo
  {volume} {11}},\ \bibinfo {pages} {1964} (\bibinfo {year}
  {2017})}\BibitemShut {NoStop}%
\bibitem [{\citenamefont {Leong}\ \emph {et~al.}(2018)\citenamefont {Leong},
  \citenamefont {Ji}, \citenamefont {Mao}, \citenamefont {Han}, \citenamefont
  {Wang}, \citenamefont {Goodman}, \citenamefont {Vignon}, \citenamefont {Su},
  \citenamefont {Guo}, \citenamefont {Shen}, \citenamefont {Gao}, \citenamefont
  {Muller}, \citenamefont {Tisdale},\ and\ \citenamefont {Kong}}]{Leong2018}%
  \BibitemOpen
  \bibfield  {author} {\bibinfo {author} {\bibfnamefont {W.~S.}\ \bibnamefont
  {Leong}}, \bibinfo {author} {\bibfnamefont {Q.}~\bibnamefont {Ji}}, \bibinfo
  {author} {\bibfnamefont {N.}~\bibnamefont {Mao}}, \bibinfo {author}
  {\bibfnamefont {Y.}~\bibnamefont {Han}}, \bibinfo {author} {\bibfnamefont
  {H.}~\bibnamefont {Wang}}, \bibinfo {author} {\bibfnamefont {A.~J.}\
  \bibnamefont {Goodman}}, \bibinfo {author} {\bibfnamefont {A.}~\bibnamefont
  {Vignon}}, \bibinfo {author} {\bibfnamefont {C.}~\bibnamefont {Su}}, \bibinfo
  {author} {\bibfnamefont {Y.}~\bibnamefont {Guo}}, \bibinfo {author}
  {\bibfnamefont {P.-C.}\ \bibnamefont {Shen}}, \bibinfo {author}
  {\bibfnamefont {Z.}~\bibnamefont {Gao}}, \bibinfo {author} {\bibfnamefont
  {D.~A.}\ \bibnamefont {Muller}}, \bibinfo {author} {\bibfnamefont {W.~A.}\
  \bibnamefont {Tisdale}}, \ and\ \bibinfo {author} {\bibfnamefont
  {J.}~\bibnamefont {Kong}},\ }\href {https://doi.org/10.1021/jacs.8b07806}
  {\bibfield  {journal} {\bibinfo  {journal} {Journal of the American Chemical
  Society}\ }\textbf {\bibinfo {volume} {140}},\ \bibinfo {pages} {12354}
  (\bibinfo {year} {2018})}\BibitemShut {NoStop}%
\bibitem [{\citenamefont {Rao}\ and\ \citenamefont {Waghmare}(2017)}]{Rao2017}%
  \BibitemOpen
  \bibfield  {author} {\bibinfo {author} {\bibfnamefont {C.~N.~R.}\
  \bibnamefont {Rao}}\ and\ \bibinfo {author} {\bibfnamefont {U.~V.}\
  \bibnamefont {Waghmare}},\ }\href
  {https://www.worldscientific.com/doi/abs/10.1142/q0078} {\emph {\bibinfo
  {title} {2D Inorganic Materials beyond Graphene}}}\ (\bibinfo  {publisher}
  {World Scientific (Europe)},\ \bibinfo {year} {2017})\BibitemShut {NoStop}%
\bibitem [{\citenamefont {Lee}\ \emph {et~al.}(2015)\citenamefont {Lee},
  \citenamefont {Cruz-Silva}, \citenamefont {Calder\'{\i}n}, \citenamefont
  {An~T~Nguyen}, \citenamefont {J~Hollander}, \citenamefont {Bersch},
  \citenamefont {E~Mallouk},\ and\ \citenamefont {A~Robinson}}]{Lee2015}%
  \BibitemOpen
  \bibfield  {author} {\bibinfo {author} {\bibfnamefont {C.-H.}\ \bibnamefont
  {Lee}}, \bibinfo {author} {\bibfnamefont {E.}~\bibnamefont {Cruz-Silva}},
  \bibinfo {author} {\bibfnamefont {L.}~\bibnamefont {Calder\'{\i}n}}, \bibinfo
  {author} {\bibfnamefont {M.}~\bibnamefont {An~T~Nguyen}}, \bibinfo {author}
  {\bibfnamefont {M.}~\bibnamefont {J~Hollander}}, \bibinfo {author}
  {\bibfnamefont {B.}~\bibnamefont {Bersch}}, \bibinfo {author} {\bibfnamefont
  {T.}~\bibnamefont {E~Mallouk}}, \ and\ \bibinfo {author} {\bibfnamefont
  {J.}~\bibnamefont {A~Robinson}},\ }\href@noop {} {\bibfield  {journal}
  {\bibinfo  {journal} {Scientific reports}\ }\textbf {\bibinfo {volume} {5}},\
  \bibinfo {pages} {10013} (\bibinfo {year} {2015})}\BibitemShut {NoStop}%
\bibitem [{\citenamefont {Perdew}\ \emph {et~al.}(1997)\citenamefont {Perdew},
  \citenamefont {Burke},\ and\ \citenamefont {Ernzerhof}}]{Perdew1997}%
  \BibitemOpen
  \bibfield  {author} {\bibinfo {author} {\bibfnamefont {J.~P.}\ \bibnamefont
  {Perdew}}, \bibinfo {author} {\bibfnamefont {K.}~\bibnamefont {Burke}}, \
  and\ \bibinfo {author} {\bibfnamefont {M.}~\bibnamefont {Ernzerhof}},\ }\href
  {https://link.aps.org/doi/10.1103/PhysRevLett.78.1396} {\bibfield  {journal}
  {\bibinfo  {journal} {Phys. Rev. Lett.}\ }\textbf {\bibinfo {volume} {78}},\
  \bibinfo {pages} {1396} (\bibinfo {year} {1997})}\BibitemShut {NoStop}%
\bibitem [{\citenamefont {Thonhauser}\ \emph {et~al.}(2015)\citenamefont
  {Thonhauser}, \citenamefont {Zuluaga}, \citenamefont {Arter}, \citenamefont
  {Berland}, \citenamefont {Schr\"oder},\ and\ \citenamefont
  {Hyldgaard}}]{Thonhauser2015}%
  \BibitemOpen
  \bibfield  {author} {\bibinfo {author} {\bibfnamefont {T.}~\bibnamefont
  {Thonhauser}}, \bibinfo {author} {\bibfnamefont {S.}~\bibnamefont {Zuluaga}},
  \bibinfo {author} {\bibfnamefont {C.~A.}\ \bibnamefont {Arter}}, \bibinfo
  {author} {\bibfnamefont {K.}~\bibnamefont {Berland}}, \bibinfo {author}
  {\bibfnamefont {E.}~\bibnamefont {Schr\"oder}}, \ and\ \bibinfo {author}
  {\bibfnamefont {P.}~\bibnamefont {Hyldgaard}},\ }\href
  {https://link.aps.org/doi/10.1103/PhysRevLett.115.136402} {\bibfield
  {journal} {\bibinfo  {journal} {Phys. Rev. Lett.}\ }\textbf {\bibinfo
  {volume} {115}},\ \bibinfo {pages} {136402} (\bibinfo {year}
  {2015})}\BibitemShut {NoStop}%
\bibitem [{\citenamefont {Thonhauser}\ \emph {et~al.}(2007)\citenamefont
  {Thonhauser}, \citenamefont {Cooper}, \citenamefont {Li}, \citenamefont
  {Puzder}, \citenamefont {Hyldgaard},\ and\ \citenamefont
  {Langreth}}]{Thonhauser2007}%
  \BibitemOpen
  \bibfield  {author} {\bibinfo {author} {\bibfnamefont {T.}~\bibnamefont
  {Thonhauser}}, \bibinfo {author} {\bibfnamefont {V.~R.}\ \bibnamefont
  {Cooper}}, \bibinfo {author} {\bibfnamefont {S.}~\bibnamefont {Li}}, \bibinfo
  {author} {\bibfnamefont {A.}~\bibnamefont {Puzder}}, \bibinfo {author}
  {\bibfnamefont {P.}~\bibnamefont {Hyldgaard}}, \ and\ \bibinfo {author}
  {\bibfnamefont {D.~C.}\ \bibnamefont {Langreth}},\ }\href
  {https://link.aps.org/doi/10.1103/PhysRevB.76.125112} {\bibfield  {journal}
  {\bibinfo  {journal} {Phys. Rev. B}\ }\textbf {\bibinfo {volume} {76}},\
  \bibinfo {pages} {125112} (\bibinfo {year} {2007})}\BibitemShut {NoStop}%
\bibitem [{\citenamefont {Berland}\ \emph {et~al.}(2015)\citenamefont
  {Berland}, \citenamefont {Cooper}, \citenamefont {Lee}, \citenamefont
  {Schr{\"o}der}, \citenamefont {Thonhauser}, \citenamefont {Hyldgaard},\ and\
  \citenamefont {Lundqvist}}]{Berland2015}%
  \BibitemOpen
  \bibfield  {author} {\bibinfo {author} {\bibfnamefont {K.}~\bibnamefont
  {Berland}}, \bibinfo {author} {\bibfnamefont {V.~R.}\ \bibnamefont {Cooper}},
  \bibinfo {author} {\bibfnamefont {K.}~\bibnamefont {Lee}}, \bibinfo {author}
  {\bibfnamefont {E.}~\bibnamefont {Schr{\"o}der}}, \bibinfo {author}
  {\bibfnamefont {T.}~\bibnamefont {Thonhauser}}, \bibinfo {author}
  {\bibfnamefont {P.}~\bibnamefont {Hyldgaard}}, \ and\ \bibinfo {author}
  {\bibfnamefont {B.~I.}\ \bibnamefont {Lundqvist}},\ }\href@noop {} {\bibfield
   {journal} {\bibinfo  {journal} {Reports on Progress in Physics}\ }\textbf
  {\bibinfo {volume} {78}},\ \bibinfo {pages} {066501} (\bibinfo {year}
  {2015})}\BibitemShut {NoStop}%
\bibitem [{\citenamefont {Langreth}\ \emph {et~al.}(2009)\citenamefont
  {Langreth}, \citenamefont {Lundqvist}, \citenamefont {Chakarova-Kack},
  \citenamefont {Cooper}, \citenamefont {Dion}, \citenamefont {Hyldgaard},
  \citenamefont {Kelkkanen}, \citenamefont {Kleis}, \citenamefont {Kong},
  \citenamefont {Li}, \citenamefont {Moses}, \citenamefont {Murray},
  \citenamefont {Puzder}, \citenamefont {Rydberg}, \citenamefont {Schrader},\
  and\ \citenamefont {Thonhauser}}]{Langreth2009}%
  \BibitemOpen
  \bibfield  {author} {\bibinfo {author} {\bibfnamefont {D.~C.}\ \bibnamefont
  {Langreth}}, \bibinfo {author} {\bibfnamefont {B.~I.}\ \bibnamefont
  {Lundqvist}}, \bibinfo {author} {\bibfnamefont {S.~D.}\ \bibnamefont
  {Chakarova-Kack}}, \bibinfo {author} {\bibfnamefont {V.~R.}\ \bibnamefont
  {Cooper}}, \bibinfo {author} {\bibfnamefont {M.}~\bibnamefont {Dion}},
  \bibinfo {author} {\bibfnamefont {P.}~\bibnamefont {Hyldgaard}}, \bibinfo
  {author} {\bibfnamefont {A.}~\bibnamefont {Kelkkanen}}, \bibinfo {author}
  {\bibfnamefont {J.}~\bibnamefont {Kleis}}, \bibinfo {author} {\bibfnamefont
  {L.}~\bibnamefont {Kong}}, \bibinfo {author} {\bibfnamefont {S.}~\bibnamefont
  {Li}}, \bibinfo {author} {\bibfnamefont {P.~G.}\ \bibnamefont {Moses}},
  \bibinfo {author} {\bibfnamefont {E.}~\bibnamefont {Murray}}, \bibinfo
  {author} {\bibfnamefont {A.}~\bibnamefont {Puzder}}, \bibinfo {author}
  {\bibfnamefont {H.}~\bibnamefont {Rydberg}}, \bibinfo {author} {\bibfnamefont
  {E.}~\bibnamefont {Schrader}}, \ and\ \bibinfo {author} {\bibfnamefont
  {T.}~\bibnamefont {Thonhauser}},\ }\href@noop {} {\bibfield  {journal}
  {\bibinfo  {journal} {Journal of Physics: Condensed Matter}\ }\textbf
  {\bibinfo {volume} {21}},\ \bibinfo {pages} {084203} (\bibinfo {year}
  {2009})}\BibitemShut {NoStop}%
\bibitem [{\citenamefont {Dion}\ \emph {et~al.}(2004)\citenamefont {Dion},
  \citenamefont {Rydberg}, \citenamefont {Schr\"oder}, \citenamefont
  {Langreth},\ and\ \citenamefont {Lundqvist}}]{Dion2004}%
  \BibitemOpen
  \bibfield  {author} {\bibinfo {author} {\bibfnamefont {M.}~\bibnamefont
  {Dion}}, \bibinfo {author} {\bibfnamefont {H.}~\bibnamefont {Rydberg}},
  \bibinfo {author} {\bibfnamefont {E.}~\bibnamefont {Schr\"oder}}, \bibinfo
  {author} {\bibfnamefont {D.~C.}\ \bibnamefont {Langreth}}, \ and\ \bibinfo
  {author} {\bibfnamefont {B.~I.}\ \bibnamefont {Lundqvist}},\ }\href
  {https://link.aps.org/doi/10.1103/PhysRevLett.92.246401} {\bibfield
  {journal} {\bibinfo  {journal} {Phys. Rev. Lett.}\ }\textbf {\bibinfo
  {volume} {92}},\ \bibinfo {pages} {246401} (\bibinfo {year}
  {2004})}\BibitemShut {NoStop}%
\bibitem [{\citenamefont {Rom\'an-P\'erez}\ and\ \citenamefont
  {Soler}(2009)}]{Roman2009}%
  \BibitemOpen
  \bibfield  {author} {\bibinfo {author} {\bibfnamefont {G.}~\bibnamefont
  {Rom\'an-P\'erez}}\ and\ \bibinfo {author} {\bibfnamefont {J.~M.}\
  \bibnamefont {Soler}},\ }\href
  {https://link.aps.org/doi/10.1103/PhysRevLett.103.096102} {\bibfield
  {journal} {\bibinfo  {journal} {Phys. Rev. Lett.}\ }\textbf {\bibinfo
  {volume} {103}},\ \bibinfo {pages} {096102} (\bibinfo {year}
  {2009})}\BibitemShut {NoStop}%
\bibitem [{\citenamefont {Joon~Choi}\ and\ \citenamefont
  {Ihm}(1999)}]{Choi1999}%
  \BibitemOpen
  \bibfield  {author} {\bibinfo {author} {\bibfnamefont {H.}~\bibnamefont
  {Joon~Choi}}\ and\ \bibinfo {author} {\bibfnamefont {J.}~\bibnamefont
  {Ihm}},\ }\href {https://link.aps.org/doi/10.1103/PhysRevB.59.2267}
  {\bibfield  {journal} {\bibinfo  {journal} {Phys. Rev. B}\ }\textbf {\bibinfo
  {volume} {59}},\ \bibinfo {pages} {2267} (\bibinfo {year}
  {1999})}\BibitemShut {NoStop}%
\bibitem [{\citenamefont {Giannozzi}\ \emph {et~al.}(2009)\citenamefont
  {Giannozzi}, \citenamefont {Baroni}, \citenamefont {Bonini}, \citenamefont
  {Calandra}, \citenamefont {Car}, \citenamefont {Cavazzoni}, \citenamefont
  {Ceresoli}, \citenamefont {Chiarotti}, \citenamefont {Cococcioni},
  \citenamefont {Dabo}, \citenamefont {Corso}, \citenamefont {de~Gironcoli},
  \citenamefont {Fabris}, \citenamefont {Fratesi}, \citenamefont {Gebauer},
  \citenamefont {Gerstmann}, \citenamefont {Gougoussis}, \citenamefont
  {Kokalj}, \citenamefont {Lazzeri}, \citenamefont {Martin-Samos},
  \citenamefont {Marzari}, \citenamefont {Mauri}, \citenamefont {Mazzarello},
  \citenamefont {Paolini}, \citenamefont {Pasquarello}, \citenamefont
  {Paulatto}, \citenamefont {Sbraccia}, \citenamefont {Scandolo}, \citenamefont
  {Sclauzero}, \citenamefont {Seitsonen}, \citenamefont {Smogunov},
  \citenamefont {Umari},\ and\ \citenamefont {Wentzcovitch}}]{Giannozzi2009}%
  \BibitemOpen
  \bibfield  {author} {\bibinfo {author} {\bibfnamefont {P.}~\bibnamefont
  {Giannozzi}}, \bibinfo {author} {\bibfnamefont {S.}~\bibnamefont {Baroni}},
  \bibinfo {author} {\bibfnamefont {N.}~\bibnamefont {Bonini}}, \bibinfo
  {author} {\bibfnamefont {M.}~\bibnamefont {Calandra}}, \bibinfo {author}
  {\bibfnamefont {R.}~\bibnamefont {Car}}, \bibinfo {author} {\bibfnamefont
  {C.}~\bibnamefont {Cavazzoni}}, \bibinfo {author} {\bibfnamefont
  {D.}~\bibnamefont {Ceresoli}}, \bibinfo {author} {\bibfnamefont {G.~L.}\
  \bibnamefont {Chiarotti}}, \bibinfo {author} {\bibfnamefont {M.}~\bibnamefont
  {Cococcioni}}, \bibinfo {author} {\bibfnamefont {I.}~\bibnamefont {Dabo}},
  \bibinfo {author} {\bibfnamefont {A.~D.}\ \bibnamefont {Corso}}, \bibinfo
  {author} {\bibfnamefont {S.}~\bibnamefont {de~Gironcoli}}, \bibinfo {author}
  {\bibfnamefont {S.}~\bibnamefont {Fabris}}, \bibinfo {author} {\bibfnamefont
  {G.}~\bibnamefont {Fratesi}}, \bibinfo {author} {\bibfnamefont
  {R.}~\bibnamefont {Gebauer}}, \bibinfo {author} {\bibfnamefont
  {U.}~\bibnamefont {Gerstmann}}, \bibinfo {author} {\bibfnamefont
  {C.}~\bibnamefont {Gougoussis}}, \bibinfo {author} {\bibfnamefont
  {A.}~\bibnamefont {Kokalj}}, \bibinfo {author} {\bibfnamefont
  {M.}~\bibnamefont {Lazzeri}}, \bibinfo {author} {\bibfnamefont
  {L.}~\bibnamefont {Martin-Samos}}, \bibinfo {author} {\bibfnamefont
  {N.}~\bibnamefont {Marzari}}, \bibinfo {author} {\bibfnamefont
  {F.}~\bibnamefont {Mauri}}, \bibinfo {author} {\bibfnamefont
  {R.}~\bibnamefont {Mazzarello}}, \bibinfo {author} {\bibfnamefont
  {S.}~\bibnamefont {Paolini}}, \bibinfo {author} {\bibfnamefont
  {A.}~\bibnamefont {Pasquarello}}, \bibinfo {author} {\bibfnamefont
  {L.}~\bibnamefont {Paulatto}}, \bibinfo {author} {\bibfnamefont
  {C.}~\bibnamefont {Sbraccia}}, \bibinfo {author} {\bibfnamefont
  {S.}~\bibnamefont {Scandolo}}, \bibinfo {author} {\bibfnamefont
  {G.}~\bibnamefont {Sclauzero}}, \bibinfo {author} {\bibfnamefont {A.~P.}\
  \bibnamefont {Seitsonen}}, \bibinfo {author} {\bibfnamefont {A.}~\bibnamefont
  {Smogunov}}, \bibinfo {author} {\bibfnamefont {P.}~\bibnamefont {Umari}}, \
  and\ \bibinfo {author} {\bibfnamefont {R.~M.}\ \bibnamefont {Wentzcovitch}},\
  }\href {http://stacks.iop.org/0953-8984/21/i=39/a=395502} {\bibfield
  {journal} {\bibinfo  {journal} {Journal of Physics: Condensed Matter}\
  }\textbf {\bibinfo {volume} {21}},\ \bibinfo {pages} {395502} (\bibinfo
  {year} {2009})}\BibitemShut {NoStop}%
\bibitem [{\citenamefont {Smogunov}\ \emph {et~al.}(2004)\citenamefont
  {Smogunov}, \citenamefont {Dal~Corso},\ and\ \citenamefont
  {Tosatti}}]{Smogunov2004}%
  \BibitemOpen
  \bibfield  {author} {\bibinfo {author} {\bibfnamefont {A.}~\bibnamefont
  {Smogunov}}, \bibinfo {author} {\bibfnamefont {A.}~\bibnamefont {Dal~Corso}},
  \ and\ \bibinfo {author} {\bibfnamefont {E.}~\bibnamefont {Tosatti}},\ }\href
  {https://link.aps.org/doi/10.1103/PhysRevB.70.045417} {\bibfield  {journal}
  {\bibinfo  {journal} {Phys. Rev. B}\ }\textbf {\bibinfo {volume} {70}},\
  \bibinfo {pages} {045417} (\bibinfo {year} {2004})}\BibitemShut {NoStop}%
\bibitem [{\citenamefont {Zhang}\ \emph {et~al.}(2016)\citenamefont {Zhang},
  \citenamefont {KC}, \citenamefont {Nie}, \citenamefont {Liang}, \citenamefont
  {Vandenberghe}, \citenamefont {Longo}, \citenamefont {Zheng}, \citenamefont
  {Kong}, \citenamefont {Hong}, \citenamefont {Wallace},\ and\ \citenamefont
  {Cho}}]{Zhang2016}%
  \BibitemOpen
  \bibfield  {author} {\bibinfo {author} {\bibfnamefont {C.}~\bibnamefont
  {Zhang}}, \bibinfo {author} {\bibfnamefont {S.}~\bibnamefont {KC}}, \bibinfo
  {author} {\bibfnamefont {Y.}~\bibnamefont {Nie}}, \bibinfo {author}
  {\bibfnamefont {C.}~\bibnamefont {Liang}}, \bibinfo {author} {\bibfnamefont
  {W.~G.}\ \bibnamefont {Vandenberghe}}, \bibinfo {author} {\bibfnamefont
  {R.~C.}\ \bibnamefont {Longo}}, \bibinfo {author} {\bibfnamefont
  {Y.}~\bibnamefont {Zheng}}, \bibinfo {author} {\bibfnamefont
  {F.}~\bibnamefont {Kong}}, \bibinfo {author} {\bibfnamefont {S.}~\bibnamefont
  {Hong}}, \bibinfo {author} {\bibfnamefont {R.~M.}\ \bibnamefont {Wallace}}, \
  and\ \bibinfo {author} {\bibfnamefont {K.}~\bibnamefont {Cho}},\ }\href
  {https://doi.org/10.1021/acsnano.6b00148} {\bibfield  {journal} {\bibinfo
  {journal} {ACS Nano}\ }\textbf {\bibinfo {volume} {10}},\ \bibinfo {pages}
  {7370} (\bibinfo {year} {2016})}\BibitemShut {NoStop}%
\bibitem [{\citenamefont {Park}\ \emph {et~al.}(2015)\citenamefont {Park},
  \citenamefont {Yun}, \citenamefont {Kim}, \citenamefont {Park}, \citenamefont
  {Chae}, \citenamefont {An}, \citenamefont {Kim}, \citenamefont {Kim},
  \citenamefont {Kim},\ and\ \citenamefont {Lee}}]{Park2015}%
  \BibitemOpen
  \bibfield  {author} {\bibinfo {author} {\bibfnamefont {J.~C.}\ \bibnamefont
  {Park}}, \bibinfo {author} {\bibfnamefont {S.~J.}\ \bibnamefont {Yun}},
  \bibinfo {author} {\bibfnamefont {H.}~\bibnamefont {Kim}}, \bibinfo {author}
  {\bibfnamefont {J.-H.}\ \bibnamefont {Park}}, \bibinfo {author}
  {\bibfnamefont {S.~H.}\ \bibnamefont {Chae}}, \bibinfo {author}
  {\bibfnamefont {S.-J.}\ \bibnamefont {An}}, \bibinfo {author} {\bibfnamefont
  {J.-G.}\ \bibnamefont {Kim}}, \bibinfo {author} {\bibfnamefont {S.~M.}\
  \bibnamefont {Kim}}, \bibinfo {author} {\bibfnamefont {K.~K.}\ \bibnamefont
  {Kim}}, \ and\ \bibinfo {author} {\bibfnamefont {Y.~H.}\ \bibnamefont
  {Lee}},\ }\href {https://doi.org/10.1021/acsnano.5b02511} {\bibfield
  {journal} {\bibinfo  {journal} {ACS Nano}\ }\textbf {\bibinfo {volume} {9}},\
  \bibinfo {pages} {6548} (\bibinfo {year} {2015})}\BibitemShut {NoStop}%
\bibitem [{\citenamefont {Sze}(1981)}]{Sze1981}%
  \BibitemOpen
  \bibfield  {author} {\bibinfo {author} {\bibfnamefont {S.}~\bibnamefont
  {Sze}},\ }\href {https://books.google.com/books?id=LCNTAAAAMAAJ} {\emph
  {\bibinfo {title} {Physics of Semiconductor Devices}}},\ Wiley-Interscience
  publication\ (\bibinfo  {publisher} {John Wiley \& Sons},\ \bibinfo {year}
  {1981})\BibitemShut {NoStop}%
\bibitem [{\citenamefont {{Padovani}}\ and\ \citenamefont
  {{Stratton}}(1966)}]{Padovani1966}%
  \BibitemOpen
  \bibfield  {author} {\bibinfo {author} {\bibfnamefont {F.~A.}\ \bibnamefont
  {{Padovani}}}\ and\ \bibinfo {author} {\bibfnamefont {R.}~\bibnamefont
  {{Stratton}}},\ }\href@noop {} {\bibfield  {journal} {\bibinfo  {journal}
  {Solid State Electronics}\ }\textbf {\bibinfo {volume} {9}},\ \bibinfo
  {pages} {695} (\bibinfo {year} {1966})}\BibitemShut {NoStop}%
\bibitem [{\citenamefont {Picozzi}\ \emph {et~al.}(2000)\citenamefont
  {Picozzi}, \citenamefont {Continenza}, \citenamefont {Satta}, \citenamefont
  {Massidda},\ and\ \citenamefont {Freeman}}]{Picozzi2000}%
  \BibitemOpen
  \bibfield  {author} {\bibinfo {author} {\bibfnamefont {S.}~\bibnamefont
  {Picozzi}}, \bibinfo {author} {\bibfnamefont {A.}~\bibnamefont {Continenza}},
  \bibinfo {author} {\bibfnamefont {G.}~\bibnamefont {Satta}}, \bibinfo
  {author} {\bibfnamefont {S.}~\bibnamefont {Massidda}}, \ and\ \bibinfo
  {author} {\bibfnamefont {A.~J.}\ \bibnamefont {Freeman}},\ }\href
  {https://link.aps.org/doi/10.1103/PhysRevB.61.16736} {\bibfield  {journal}
  {\bibinfo  {journal} {Phys. Rev. B}\ }\textbf {\bibinfo {volume} {61}},\
  \bibinfo {pages} {16736} (\bibinfo {year} {2000})}\BibitemShut {NoStop}%
\bibitem [{\citenamefont {B\"uttiker}\ and\ \citenamefont
  {Landauer}(1982)}]{buttiker1982}%
  \BibitemOpen
  \bibfield  {author} {\bibinfo {author} {\bibfnamefont {M.}~\bibnamefont
  {B\"uttiker}}\ and\ \bibinfo {author} {\bibfnamefont {R.}~\bibnamefont
  {Landauer}},\ }\href {https://link.aps.org/doi/10.1103/PhysRevLett.49.1739}
  {\bibfield  {journal} {\bibinfo  {journal} {Phys. Rev. Lett.}\ }\textbf
  {\bibinfo {volume} {49}},\ \bibinfo {pages} {1739} (\bibinfo {year}
  {1982})}\BibitemShut {NoStop}%
\bibitem [{\citenamefont {Datta}(1995)}]{datta1995}%
  \BibitemOpen
  \bibfield  {author} {\bibinfo {author} {\bibfnamefont {S.}~\bibnamefont
  {Datta}},\ }\href@noop {} {\emph {\bibinfo {title} {Electronic Transport in
  Mesoscopic Systems}}},\ Cambridge Studies in Semiconductor Physics and
  Microelectronic Engineering\ (\bibinfo  {publisher} {Cambridge University
  Press},\ \bibinfo {year} {1995})\BibitemShut {NoStop}%
\bibitem [{\citenamefont {Di~Carlo}\ \emph {et~al.}(1994)\citenamefont
  {Di~Carlo}, \citenamefont {Vogl},\ and\ \citenamefont {P\"otz}}]{Carlo1994}%
  \BibitemOpen
  \bibfield  {author} {\bibinfo {author} {\bibfnamefont {A.}~\bibnamefont
  {Di~Carlo}}, \bibinfo {author} {\bibfnamefont {P.}~\bibnamefont {Vogl}}, \
  and\ \bibinfo {author} {\bibfnamefont {W.}~\bibnamefont {P\"otz}},\ }\href
  {https://link.aps.org/doi/10.1103/PhysRevB.50.8358} {\bibfield  {journal}
  {\bibinfo  {journal} {Phys. Rev. B}\ }\textbf {\bibinfo {volume} {50}},\
  \bibinfo {pages} {8358} (\bibinfo {year} {1994})}\BibitemShut {NoStop}%
\bibitem [{\citenamefont {Kuroda}\ \emph {et~al.}(2011)\citenamefont {Kuroda},
  \citenamefont {Tersoff}, \citenamefont {Newns},\ and\ \citenamefont
  {Martyna}}]{kuroda2011}%
  \BibitemOpen
  \bibfield  {author} {\bibinfo {author} {\bibfnamefont {M.~A.}\ \bibnamefont
  {Kuroda}}, \bibinfo {author} {\bibfnamefont {J.}~\bibnamefont {Tersoff}},
  \bibinfo {author} {\bibfnamefont {D.~M.}\ \bibnamefont {Newns}}, \ and\
  \bibinfo {author} {\bibfnamefont {G.~J.}\ \bibnamefont {Martyna}},\
  }\href@noop {} {\bibfield  {journal} {\bibinfo  {journal} {Nano Lett.}\
  }\textbf {\bibinfo {volume} {11}},\ \bibinfo {pages} {3629} (\bibinfo {year}
  {2011})}\BibitemShut {NoStop}%
\bibitem [{\citenamefont {Chuang}\ \emph {et~al.}(2016)\citenamefont {Chuang},
  \citenamefont {Chamlagain}, \citenamefont {Koehler}, \citenamefont {Perera},
  \citenamefont {Yan}, \citenamefont {Mandrus}, \citenamefont {Tománek},\ and\
  \citenamefont {Zhou}}]{Chuang2016}%
  \BibitemOpen
  \bibfield  {author} {\bibinfo {author} {\bibfnamefont {H.-J.}\ \bibnamefont
  {Chuang}}, \bibinfo {author} {\bibfnamefont {B.}~\bibnamefont {Chamlagain}},
  \bibinfo {author} {\bibfnamefont {M.}~\bibnamefont {Koehler}}, \bibinfo
  {author} {\bibfnamefont {M.~M.}\ \bibnamefont {Perera}}, \bibinfo {author}
  {\bibfnamefont {J.}~\bibnamefont {Yan}}, \bibinfo {author} {\bibfnamefont
  {D.}~\bibnamefont {Mandrus}}, \bibinfo {author} {\bibfnamefont
  {D.}~\bibnamefont {Tománek}}, \ and\ \bibinfo {author} {\bibfnamefont
  {Z.}~\bibnamefont {Zhou}},\ }\href
  {https://doi.org/10.1021/acs.nanolett.5b05066} {\bibfield  {journal}
  {\bibinfo  {journal} {Nano Letters}\ }\textbf {\bibinfo {volume} {16}},\
  \bibinfo {pages} {1896} (\bibinfo {year} {2016})}\BibitemShut {NoStop}%
\bibitem [{\citenamefont {Zhang}\ \emph {et~al.}(2014)\citenamefont {Zhang},
  \citenamefont {Johnson}, \citenamefont {Hsu}, \citenamefont {Li},\ and\
  \citenamefont {Shih}}]{zhang2014a}%
  \BibitemOpen
  \bibfield  {author} {\bibinfo {author} {\bibfnamefont {C.}~\bibnamefont
  {Zhang}}, \bibinfo {author} {\bibfnamefont {A.}~\bibnamefont {Johnson}},
  \bibinfo {author} {\bibfnamefont {C.-L.}\ \bibnamefont {Hsu}}, \bibinfo
  {author} {\bibfnamefont {L.-J.}\ \bibnamefont {Li}}, \ and\ \bibinfo {author}
  {\bibfnamefont {C.-K.}\ \bibnamefont {Shih}},\ }\bibfield  {booktitle} {\emph
  {\bibinfo {booktitle} {Nano Letters}},\ }\href
  {https://doi.org/10.1021/nl501133c} {\bibfield  {journal} {\bibinfo
  {journal} {Nano Letters}\ }\textbf {\bibinfo {volume} {14}},\ \bibinfo
  {pages} {2443} (\bibinfo {year} {2014})}\BibitemShut {NoStop}%
\bibitem [{\citenamefont {Tosun}\ \emph {et~al.}(2016)\citenamefont {Tosun},
  \citenamefont {Chan}, \citenamefont {Amani}, \citenamefont {Roy},
  \citenamefont {Ahn}, \citenamefont {Taheri}, \citenamefont {Carraro},
  \citenamefont {Ager}, \citenamefont {Maboudian},\ and\ \citenamefont
  {Javey}}]{Tosun2016}%
  \BibitemOpen
  \bibfield  {author} {\bibinfo {author} {\bibfnamefont {M.}~\bibnamefont
  {Tosun}}, \bibinfo {author} {\bibfnamefont {L.}~\bibnamefont {Chan}},
  \bibinfo {author} {\bibfnamefont {M.}~\bibnamefont {Amani}}, \bibinfo
  {author} {\bibfnamefont {T.}~\bibnamefont {Roy}}, \bibinfo {author}
  {\bibfnamefont {G.~H.}\ \bibnamefont {Ahn}}, \bibinfo {author} {\bibfnamefont
  {P.}~\bibnamefont {Taheri}}, \bibinfo {author} {\bibfnamefont
  {C.}~\bibnamefont {Carraro}}, \bibinfo {author} {\bibfnamefont {J.~W.}\
  \bibnamefont {Ager}}, \bibinfo {author} {\bibfnamefont {R.}~\bibnamefont
  {Maboudian}}, \ and\ \bibinfo {author} {\bibfnamefont {A.}~\bibnamefont
  {Javey}},\ }\href {https://doi.org/10.1021/acsnano.6b02521} {\bibfield
  {journal} {\bibinfo  {journal} {ACS Nano}\ }\textbf {\bibinfo {volume}
  {10}},\ \bibinfo {pages} {6853} (\bibinfo {year} {2016})}\BibitemShut
  {NoStop}%
\bibitem [{\citenamefont {Xu}\ \emph {et~al.}(2017)\citenamefont {Xu},
  \citenamefont {Wang}, \citenamefont {Zhao},\ and\ \citenamefont
  {Chai}}]{Xu2017}%
  \BibitemOpen
  \bibfield  {author} {\bibinfo {author} {\bibfnamefont {K.}~\bibnamefont
  {Xu}}, \bibinfo {author} {\bibfnamefont {Y.}~\bibnamefont {Wang}}, \bibinfo
  {author} {\bibfnamefont {Y.}~\bibnamefont {Zhao}}, \ and\ \bibinfo {author}
  {\bibfnamefont {Y.}~\bibnamefont {Chai}},\ }\href {\doibase
  10.1039/C6TC04640A} {\bibfield  {journal} {\bibinfo  {journal} {J. Mater.
  Chem. C}\ }\textbf {\bibinfo {volume} {5}},\ \bibinfo {pages} {376} (\bibinfo
  {year} {2017})}\BibitemShut {NoStop}%
\bibitem [{\citenamefont {{Marian}}\ \emph {et~al.}(2016)\citenamefont
  {{Marian}}, \citenamefont {{Dib}}, \citenamefont {{Cusati}}, \citenamefont
  {{Fortunelli}}, \citenamefont {{Iannaccone}},\ and\ \citenamefont
  {{Fiori}}}]{marian2016}%
  \BibitemOpen
  \bibfield  {author} {\bibinfo {author} {\bibfnamefont {D.}~\bibnamefont
  {{Marian}}}, \bibinfo {author} {\bibfnamefont {E.}~\bibnamefont {{Dib}}},
  \bibinfo {author} {\bibfnamefont {T.}~\bibnamefont {{Cusati}}}, \bibinfo
  {author} {\bibfnamefont {A.}~\bibnamefont {{Fortunelli}}}, \bibinfo {author}
  {\bibfnamefont {G.}~\bibnamefont {{Iannaccone}}}, \ and\ \bibinfo {author}
  {\bibfnamefont {G.}~\bibnamefont {{Fiori}}},\ }\href@noop {} {\bibfield
  {journal} {\bibinfo  {journal} {2016 IEEE IEDM}\ ,\ \bibinfo {pages}
  {14.1.1}} (\bibinfo {year} {2016})}\BibitemShut {NoStop}%
\bibitem [{\citenamefont {M.~Ugeda}\ \emph {et~al.}(2018)\citenamefont
  {M.~Ugeda}, \citenamefont {Pulkin}, \citenamefont {Tang}, \citenamefont
  {Ryu}, \citenamefont {Wu}, \citenamefont {Zhang}, \citenamefont {Wong},
  \citenamefont {Pedramrazi}, \citenamefont {Mart\'in-Recio}, \citenamefont
  {Chen}, \citenamefont {Wang}, \citenamefont {Shen}, \citenamefont {Mo},
  \citenamefont {Yazyev},\ and\ \citenamefont {F.~Crommie}}]{Ugeda2018}%
  \BibitemOpen
  \bibfield  {author} {\bibinfo {author} {\bibfnamefont {M.}~\bibnamefont
  {M.~Ugeda}}, \bibinfo {author} {\bibfnamefont {A.}~\bibnamefont {Pulkin}},
  \bibinfo {author} {\bibfnamefont {S.}~\bibnamefont {Tang}}, \bibinfo {author}
  {\bibfnamefont {H.}~\bibnamefont {Ryu}}, \bibinfo {author} {\bibfnamefont
  {Q.-S.}\ \bibnamefont {Wu}}, \bibinfo {author} {\bibfnamefont
  {Y.}~\bibnamefont {Zhang}}, \bibinfo {author} {\bibfnamefont
  {D.}~\bibnamefont {Wong}}, \bibinfo {author} {\bibfnamefont {Z.}~\bibnamefont
  {Pedramrazi}}, \bibinfo {author} {\bibfnamefont {A.}~\bibnamefont
  {Mart\'in-Recio}}, \bibinfo {author} {\bibfnamefont {Y.}~\bibnamefont
  {Chen}}, \bibinfo {author} {\bibfnamefont {F.}~\bibnamefont {Wang}}, \bibinfo
  {author} {\bibfnamefont {Z.-X.}\ \bibnamefont {Shen}}, \bibinfo {author}
  {\bibfnamefont {S.-K.}\ \bibnamefont {Mo}}, \bibinfo {author} {\bibfnamefont
  {O.}~\bibnamefont {Yazyev}}, \ and\ \bibinfo {author} {\bibfnamefont
  {M.}~\bibnamefont {F.~Crommie}},\ }\href@noop {} {\bibfield  {journal}
  {\bibinfo  {journal} {Nature Communications}\ }\textbf {\bibinfo {volume}
  {9}},\ \bibinfo {pages} {3401} (\bibinfo {year} {2018})}\BibitemShut
  {NoStop}%
\bibitem [{\citenamefont {Wang}\ \emph {et~al.}(2019)\citenamefont {Wang},
  \citenamefont {Li}, \citenamefont {Chen}, \citenamefont {Deng},\ and\
  \citenamefont {Niu}}]{Wang2019}%
  \BibitemOpen
  \bibfield  {author} {\bibinfo {author} {\bibfnamefont {J.}~\bibnamefont
  {Wang}}, \bibinfo {author} {\bibfnamefont {Z.}~\bibnamefont {Li}}, \bibinfo
  {author} {\bibfnamefont {H.}~\bibnamefont {Chen}}, \bibinfo {author}
  {\bibfnamefont {G.}~\bibnamefont {Deng}}, \ and\ \bibinfo {author}
  {\bibfnamefont {X.}~\bibnamefont {Niu}},\ }\href
  {https://doi.org/10.1007/s40820-019-0276-y} {\bibfield  {journal} {\bibinfo
  {journal} {Nano-Micro Letters}\ }\textbf {\bibinfo {volume} {11}},\ \bibinfo
  {pages} {48} (\bibinfo {year} {2019})}\BibitemShut {NoStop}%
\bibitem [{\citenamefont {Chen}\ \emph {et~al.}(2017)\citenamefont {Chen},
  \citenamefont {Yang}, \citenamefont {Zhang},\ and\ \citenamefont
  {Kaxiras}}]{Chen2017}%
  \BibitemOpen
  \bibfield  {author} {\bibinfo {author} {\bibfnamefont {W.}~\bibnamefont
  {Chen}}, \bibinfo {author} {\bibfnamefont {Y.}~\bibnamefont {Yang}}, \bibinfo
  {author} {\bibfnamefont {Z.}~\bibnamefont {Zhang}}, \ and\ \bibinfo {author}
  {\bibfnamefont {E.}~\bibnamefont {Kaxiras}},\ }\href@noop {} {\bibfield
  {journal} {\bibinfo  {journal} {2D Materials}\ }\textbf {\bibinfo {volume}
  {4}},\ \bibinfo {pages} {045001} (\bibinfo {year} {2017})}\BibitemShut
  {NoStop}%
\bibitem [{\citenamefont {Roth}\ \emph {et~al.}(2009)\citenamefont {Roth},
  \citenamefont {Br{\"u}ne}, \citenamefont {Buhmann}, \citenamefont
  {Molenkamp}, \citenamefont {Maciejko}, \citenamefont {Qi},\ and\
  \citenamefont {Zhang}}]{roth2009}%
  \BibitemOpen
  \bibfield  {author} {\bibinfo {author} {\bibfnamefont {A.}~\bibnamefont
  {Roth}}, \bibinfo {author} {\bibfnamefont {C.}~\bibnamefont {Br{\"u}ne}},
  \bibinfo {author} {\bibfnamefont {H.}~\bibnamefont {Buhmann}}, \bibinfo
  {author} {\bibfnamefont {L.~W.}\ \bibnamefont {Molenkamp}}, \bibinfo {author}
  {\bibfnamefont {J.}~\bibnamefont {Maciejko}}, \bibinfo {author}
  {\bibfnamefont {X.-L.}\ \bibnamefont {Qi}}, \ and\ \bibinfo {author}
  {\bibfnamefont {S.-C.}\ \bibnamefont {Zhang}},\ }\href
  {http://science.sciencemag.org/content/325/5938/294} {\bibfield  {journal}
  {\bibinfo  {journal} {Science}\ }\textbf {\bibinfo {volume} {325}},\ \bibinfo
  {pages} {294} (\bibinfo {year} {2009})}\BibitemShut {NoStop}%
\bibitem [{\citenamefont {\'Avalos-Ovando}\ \emph {et~al.}(2019)\citenamefont
  {\'Avalos-Ovando}, \citenamefont {Mastrogiuseppe},\ and\ \citenamefont
  {Ulloa}}]{Ovando2019}%
  \BibitemOpen
  \bibfield  {author} {\bibinfo {author} {\bibfnamefont {O.}~\bibnamefont
  {\'Avalos-Ovando}}, \bibinfo {author} {\bibfnamefont {D.}~\bibnamefont
  {Mastrogiuseppe}}, \ and\ \bibinfo {author} {\bibfnamefont {S.~E.}\
  \bibnamefont {Ulloa}},\ }\href
  {https://link.aps.org/doi/10.1103/PhysRevB.99.035107} {\bibfield  {journal}
  {\bibinfo  {journal} {Phys. Rev. B}\ }\textbf {\bibinfo {volume} {99}},\
  \bibinfo {pages} {035107} (\bibinfo {year} {2019})}\BibitemShut {NoStop}%
\end{thebibliography}%

\end{document}